\documentclass{article}

\pdfoutput=1
\usepackage{arxiv}

\usepackage[utf8]{inputenc} 
\usepackage[T1]{fontenc}    
\usepackage{hyperref}       
\usepackage{url}            
\usepackage{booktabs}       
\usepackage{amsfonts}       
\usepackage{nicefrac}       
\usepackage{microtype}      
\usepackage{lipsum}
\usepackage[nolist]{acronym}
\usepackage{amsmath, amsthm, amssymb, amsfonts}
\usepackage{enumitem}

\usepackage{tabularx}
\usepackage{fancyhdr}
\usepackage{fancybox}
\usepackage{amsmath, amsthm, amssymb, amsfonts}
\usepackage{tabularx, ragged2e, booktabs, caption}
\usepackage{subfig}
\usepackage{longtable}
\usepackage{xfrac}
\usepackage{enumitem}	
\usepackage{afterpage}
\usepackage[utf8]{inputenc}
\usepackage{hyperref}
\usepackage{epstopdf}
\usepackage{color}

\title{Plant-wide fault and disturbance screening using combined transfer entropy and eigenvector centrality analysis}

\author{
  Simon J.~Streicher \\
  Department of Chemical Engineering\\
  University of Pretoria\\
  South Africa \\
   \And
 Carl~Sandrock \\
  Department of Chemical Engineering\\
  University of Pretoria\\
  South Africa \\
  \texttt{carl.sandrock@up.ac.za} \\
}

\begin{document}
\maketitle

\begin{acronym}
\acro{AIS} {Active Information Storage}
\acro{APC}{advanced process control}
\acro{APM}{advanced process monitoring}
\acro{CAEX}{Computer Aided Engineering Exchange}
\acro{CC}{Cross Correlation}
\acro{CCA}{Canonical Correlation Analysis}
\acro{CCM}{Cross Convergent Mapping}
\acro{CLEB}{Control Loop Economic Benefit}
\acro{CMI}{Conditional Mutual Information}
\acro{CPI}{Controller Performance Index}
\acro{CV}{Controlled Variable}
\acro{DAG}{Directed Ascyclic Graph}
\acro{DTE}{Directed Transfer Entropy}
\acro{DV}{Disturbance Variable}
\acro{EWMA}{Exponentially Weighted Moving Average}
\acro{EKF}{Extended Kalmal Filter}
\acro{FDA}{Fischer Discriminant Analysis}
\acro{FDD}{fault detection and diagnosis}
\acro{FFT}{fast Fourier transform}
\acro{FOPDT}{First Order Plus Dead Time}
\acro{GC}{Ganger Causality}
\acro{HPC}{High Performance Computing}
\acro{iAAFT}{iterative amplitude adjusted Fourier transform}
\acro{IAE}{Integrated Absolute Error}
\acro{ISE}{Integrated Square of Error}
\acro{ICA}{Independent Component Analysis}
\acro{ITD}{Interaction in Time Domain}
\acro{ITN}{information transfer network}
\acro{JIDT}{Java Information Dynamics Toolkit}
\acro{KPI}{key performance indicator}
\acro{$k$-NN}{$k$ - nearest neighbour}
\acro{KSG}{Kraskov-St{\"u}gbauer-Grassberger}
\acro{LDDP}{Limiting Density of Discrete Points}
\acro{MPC}{Model Predictive Control}
\acro{MTR}{multiple time region}
\acro{MV}{manipulated variable}
\acro{PCA}{principal component analysis}
\acro{PDF}{probability density function}
\acro{PFD}{process flow diagram}
\acro{PID}{Proportional Integral Derivative}
\acro{PnID}[P\&ID]{Pipeline and Instrumentation Diagram}
\acro{PLS}{Partial Least Squares}
\acro{PGM}{Probablistic Graphical Model}
\acro{PSCI}{Power Spectral Correlation Index}
\acro{PTE}{Partial Transfer Entropy}
\acro{PV}{Process Variable}
\acro{QTA}{Qualitative Trend Analysis}
\acro{RBC}{reconstruction based components}
\acro{RGA}{Relative Gain Array}
\acro{RPM}{reset probability matrix}
\acro{SDG}{signed directed graph}
\acro{SPC}{Statistical Process Control}
\acro{SVD}{Singular Value Decomposition}
\acro{SVM}{Support Vector Machines}
\acro{TE}{transfer entropy}
\acro{TPM}{transistion probability matrix}
\acro{XML}{Extensible Markup Language}
\end{acronym}

\begin{abstract}
	Finding the source of a disturbance or fault in complex systems such as industrial chemical processing plants can be a difficult task and consume a significant number of engineering hours.
	In many cases, a systematic elimination procedure is considered to be the only feasible approach but can cause undesired process upsets.
	Practitioners desire robust alternative approaches.
	
	This paper presents an unsupervised, data-driven method for ranking process elements according to the magnitude and novelty of their influence.
	Partial bivariate transfer entropy estimation is used to infer a weighted directed graph of process elements.
	Eigenvector centrality is applied to rank network nodes according to their overall effect.
	As the ranking of process elements rely on emerging properties that depends on the aggregate of many connections, the results are robust to errors in the estimation of individual edge properties and the inclusion of indirect connections that do not represent the true causal structure of the process.
	
	A monitoring chart of continuously calculated process element importance scores over multiple overlapping time regions can assist with incipient fault detection.
	Ranking results combined with visual inspection of information transfer networks is also useful for root cause analysis of known faults and disturbances. A software implementation of the proposed method is available.
\end{abstract}


\section{Introduction}

Plant-wide \ac{FDD} and \ac{APM} is an active topic in control engineering research, and has been for many decades \cite{Thornhill2006, Venkatasubramanian2003}.
However, many proposed methods require either a database of known faults or a benchmark of nominal, fault-free operation.
These requirements are a significant barrier to entry for widespread industrial adoption which requires methods to be robust and easy to implement and maintain.

An optimal \ac{APM} system will assess all process states in near real-time, drawing attention to the most critical behaviour, provide insight into the causal flow of events and offer actionable advisories such that operating personnel can take optimal corrective remedial action.
Ranking faults and poor control loop performance based on overall impact and importance is a useful triaging strategy for over-extended control departments ~\cite{Farenzena2009a, Farenzena2009b, Rahman2011}.
A good ranking tool will rank faults and disturbances such that those affecting the most important variables in the system is assigned higher priority than conditions which may be more severe but affecting less important variables.
The flagged areas should be as close to the real source as sensor placement allows, even if significantly amplified by other processes or control loops.

\section{Research framework}
The time-series signals captured from periodic sampling of measurement devices from  a chemical process under automatic closed-loop control can be expected to show the behaviour of a continuous, dynamic multivariate system that is complex, non-linear and non-stationary.
Systems with these properties are relatively challenging to analyse compared to discrete, stationary, univariate and linear systems, for which the theory is well developed.

Chemical processes are inherently deterministic.
Every single measurement reflected by every probe and each floating point value stored in a process historian is ultimately the result of a deterministic physical system that follows a set of heavily convoluted but individually simple laws.
Fundamental processes that transfer heat and mass between elements and chemical reactions that alter the composition of streams do not generate original information.
If the entropy of states exciting a node is more than the incoming entropy, the additional information must have originated from a disturbance or fault.
This property is not true for some other complex systems frequently studied, such as a stock market.
The proposed method for \ac{FDD} presented here relies on the premise that uncertainty introduced into a system that impacts significant variables indicates the severity and location of a fault or disturbance, and that a combination of information and graph theory can trace this uncertainty back to its original source.

\section{Prior art and context of proposed methods}
\subsection{Use of directed graphs in \ac{FDD}}

\begin{figure}[h]
	\begin{minipage}[c]{0.60\linewidth}
		\centering
		\includegraphics[scale=0.8]{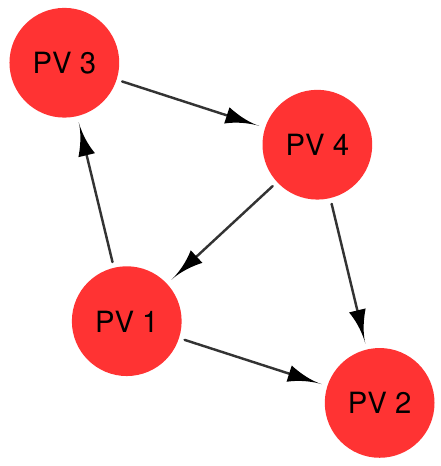}
	\end{minipage}%
	\begin{minipage}[c]{0.30\linewidth}
		\centering
		\[ \mathbf{A} = \left[\begin{array}{llll}
		0 & 0 & 0 & 1\\
		1 & 0 & 0& 1\\
		1 & 0 & 0 & 0\\
		0 & 0 & 1 & 0\end{array} \right]\]
	\end{minipage}
	\caption{ Example directed graph with associated adjacency matrix}
	\label{fig:example_digraph}
\end{figure}

A directed graph can be represented in matrix form as an adjacency matrix,
\begin{equation}\label{eq:adjacency_definition}
A_{ij} =
\begin{cases}
1, & \text{if there is a connection from node } j \text{ directed towards node } i \\
0, & \text{otherwise}
\end{cases}
\end{equation}
as indicated in Figure~\ref{fig:example_digraph}.
This work follows the convention of sources along columns and sinks along rows.
In a weighted digraph, the Boolean entries of the adjacency matrix are replaced by the appropriate weights.

The first known use of a \ac{SDG} for modelling chemical processes dates almost 40 years back \cite{Iri1979}, and was later extended to include weights based on sensitivity coefficients \cite{Vianna1995}.
Data-driven methods and heuristics captured from operational experience can be used for their development, and a complete quantitative description is not required \cite{Maurya2006}.
\ac{SDG} models provide a representation that captures cause and effect relationships which can be easily applied in diagnostic reasoning, making them a popular component of several proposed approaches to plant-wide \ac{FDD} \cite{Yang2013, Yang2014}.

\subsection{Transfer entropy based causal inference}

The idea of inferring an \ac{ITN} using the \ac{TE} functional for \ac{FDD} applications has gathered significant traction among several research groups over the last decade.
\ac{TE} relates to other information theoretic measures as follows: 
The average \emph{entropy} of a signal is a measure of its uncertainty;
observations of uncertain states carry more information than samples from predictable signals.
\emph{Conditional entropy} gives the remaining entropy after accounting for uncertainty introduced due to the effect of other processes.
\emph{Mutual information} measures entropy not unique to a specific signal but shared with other signals evaluated, while \emph{conditional mutual information} is the remaining entropy shared among a set of signals after accounting for the effect of other processes.
\emph{Transfer entropy} is the entropy shared between a source and target signal after conditioning on the target signal's past.
It can be viewed as a measure of conditional mutual information in time.
Finally, \emph{total \ac{TE}} is the entropy \emph{uniquely} shared between a source and target signal measured by conditioning on every other signal in the system under investigation in addition to the target's past \cite{Cover2006, Faes2016}.

Most proposed methods involving \ac{TE} in the context of \ac{FDD} try to develop an \ac{ITN} for the express purpose of reasoning based on its structure \cite{Bauer2005a, Landman2014, Yang2014}.
Significant effort has been made to avoid spurious or indirect connections from being included \cite{Duan2013} and to reduce the computational load by symbolic \cite{Staniek2008} and discrete approximations of signals based on alarm limits \cite{Yu2015, Su2017a}.
In contrast, the method proposed here relies on emerging properties that depend on the aggregate of many connections, and as such is more robust to errors in the estimation of individual causal links.
In fact, including indirect connections between source and target nodes will boost the importance of influential nodes, making the detection of changes in the plant's causal structure easier.

Even simplified models of a typical chemical plant are likely to involve several hundreds of states.
Rigorous data-driven information decomposition for a general non-linear process becomes computationally intractable past a few states, while analytical solutions for multivariate Gaussian processes exist \cite{Faes2017a}.
As a second reward to full multivariate information decomposition, \emph{complete} bivariate \ac{TE} analysis that accounts for the effects of all potential sources can provide a sufficient approximation of an \ac{ITN} to allow for causal reasoning \cite{Runge2012, Duan2015a}.
Estimating complete \ac{TE} still suffers from the curse of dimensionality, but graph theory based simplifications \cite{Runge2012} and greedy-algorithms \cite{Lizier2012} proved successful in truncating the problem to a manageable number of dimensions.  

Simplifying things further, a \emph{partial} bivariate information transfer metric combined with graph centrality measures can indicate the most important sources of original information, without necessarily providing an accurate representation of the system's unique and direct internal information pathways.
As the intended goal of the proposed method is to be a useful plant-wide fault triaging indicator and not provide causal reasoning support \emph{per se}, this approach that is two steps (partial and bivariate) away from the near-unattainable theoretical optimum of full information decomposition provides an appropriate balance between rigour and computational complexity.

\subsection{Influence based ranking of process elements}

Farenzena \cite{Farenzena2009a} suggested prioritising control loop maintenance according to a variability matrix that indicates how improving the performance of a specific loop to that of a minimum variance controller is expected to influence the variability in other control loops.
In many cases, increased performance in one section of a plant will offset disturbances to interacting units, and solving the ranking problem will give the best compromise while being insensitive to model mismatch.

The idea of using a variant of Google's PageRank-type algorithm for analysing connectivity between control loops was introduced in 2009 and dubbed LoopRank \cite{Farenzena2009b}.
Partial correlation was used to determine a weighted directed graph.
Recently a combination of transfer entropy based causal inference and eigenvector network centrality with additional post-processing steps was successfully applied to detect epilepsy seizure onset zones from neurological data \cite{Murin2018}.
The nature of this problem is similar to the task of detecting the source of an incipient disturbance in a chemical plant.

\section{Information transfer estimation}

\subsection{Formal definitions}
Transfer entropy was introduced almost two decades ago as a dynamic, non-linear and non-parametric measure for measuring information flow \cite{Schreiber2000b}:
\begin{equation}\label{schreiber_definition}
T_{Y \rightarrow X} =   \sum_{\left(x_{i + h}, \mathbf{x}_{i}^k, \mathbf{y}^l_i\right)} p \left(x_{i + 1}, \mathbf{x}_i^k, \mathbf{y}^l_i\right)\log{\frac{p \left(x_{i + 1} | \mathbf{x}_i^k, \mathbf{y}^l_i \right )}{p \left(x_{i + 1} | \mathbf{x}_i^k \right )}}
\end{equation}
where $Y$ and $X$ are the source and target variables with embedding dimensions of $l$ and $k$ respectively.
Using the natural logarithm will return values in the units of nats while using base $2$ will return bits. 

The need to optimise the measure over a range of potential lags was identified when \ac{TE} was applied to process engineering signals \cite{Bauer2005a}, resulting in the following formulation \cite{Shu2013} that includes a lag parameter $h$:
\begin{equation}\label{eq:shu_definition}
T_{Y \rightarrow X} =  \sum_{\left(x_{i + h}, \mathbf{x}_{i + h - 1}^k, \mathbf{y}^l_i\right)} p \left(x_{i + h}, \mathbf{x}_{i + h - 1}^k, \mathbf{y}^l_i\right)\log{\frac{p \left(x_{i + h} | \mathbf{x}_{i + h - 1}^k, \mathbf{y}^l_i \right )}{p \left(x_{i + h} | \mathbf{x}_{i + h - 1}^k \right )}}.
\end{equation}

As calculation will involve a finite number of observations, the estimator can be re-formulating as a global average of local transfer entropies \cite{Lizier2008}, eliminating one \ac{PDF} estimation and significantly reducing the number of states that need to be summed:
\begin{equation}\label{eq:lizier_definition}
T_{Y \rightarrow X} = \frac{1}{N} \sum_{i = 1}^{N} \log {\frac{p \left(x_{i + h} | \mathbf{x}_{i + h - 1}^k, \mathbf{y}^l_i \right )}{p \left(x_{i + h} | \mathbf{x}_{i + h - 1}^k \right )}}.
\end{equation}

Some applications in prior art used the difference between forward and backwards transfer entropy estimates to maximise the potential benefit obtained from the measure's asymmetrical properties \cite{Bauer2005a}.
The term directional as opposed to simple \ac{TE} is used to refer to this formulation in the scope of this work:
\begin{equation}\label{eq:directional_transfer_entropy}
t_{Y \rightarrow X} =  T_{Y\rightarrow X} -  T_{X\rightarrow Y}.
\end{equation}

\subsection{Estimators}
Since \ac{TE} is a functional of multi-dimensional and sometimes difficult to estimate \ac{PDF}s, the method used for \ac{PDF} approximation can have a significant influence on its accuracy.
Discrete signals with countably limited options allow for easy computation of precise \ac{PDF}s.
However, for continuous variables not modelled according to a specific statistical distribution with an analytical solution for entropy, the bandwidth and/or other parameters of the estimator will influence results.
The application of parameter-free zero information calculation to \ac{TE} provides a robust estimator applicable to non-stationary processes but tends to be overly conservative with a high false negative rate \cite{Duan2015a}.

The simplest approach to estimating \ac{TE} on continuous signals is to assume that the signal distributions are Gaussian, resulting in a simple closed-form expression.
A mean and covariance function completely specifies Gaussian processes and can be used to describe time series data from various sources with reasonable success.
However, approaching a non-Gaussian distribution as a Gaussian can result in significant entropy estimation errors \cite{Hlavackovaschindler2007}.
Numerous alternatives to Gaussian kernels are available, and are typically hand-picked according to the specific application and nature of the data \cite{Duvenaud2014}.
There are no automated routines for estimating the best kernel choice for non-Gaussian distributions, making the use of individually tuned kernels impractical as \ac{FDD} methods should be applicable to a variety and mix of signal distributions.

The \ac{JIDT} contains routines for box (or uniform) kernel estimation together with dynamic correlation exclusion while optimising box counting for computing differential \ac{TE} \cite{Lizier2008}.
Box kernels make no assumptions about the nature of distributions and are therefore robust to a wide variety of datasets from different sources at the cost of requiring more samples to achieve comparable accuracy as a shaped kernel that does correctly represent the distribution.
Box kernels are discontinuous with a square shape and are similar to histograms in univariate cases, producing non-smooth results.
The measure centres a box over each data point, and the overlap of boxes determine the density estimate at each point.
The box size also referred to as the bandwidth, is the sole parameter of this estimator and affects the level of fidelity.

The \ac{$k$-NN} based \ac{KSG} estimator \cite{Kraskov2004} is considered the current best-of-breed approach for empirical (conditional) mutual information \cite{Wibral2014} and is recommended for \ac{TE} estimation \cite{Lizier2014}.
It estimates mutual information without explicitly estimating a \ac{PDF} as an intermediate step or assuming a specific distribution.
The number of neighbours is the sole parameter and the measure is pleasantly insensitive to it.
A default value of $k = 4$ is recommended.
This measure is also available as part of \ac{JIDT}.

The use of both the box-kernel and \ac{KSG} estimators was evaluated on test cases, and the estimated magnitudes differed significantly, with \ac{KSG} estimates results exceeding significance thresholds by a larger margin.
Additionally, the \ac{KSG} directional results are very similar to that of simple estimates, indicating that this estimator correctly assigns a near-zero value to the inverse relationship.

\subsection{Embeddings}

In the most naive implementations, the embedding dimensions $l$ and $k$ are both set to unity.
Ideally, to separate information storage from transfer, the embedding dimensions of the target signal $k$ needs to be optimised such that the self-information between $x_{i+1}$ and $\mathbf{x}_i^k$ is maximised \cite{Lizier2014}.
Failure to do so will lead to an over-estimation of \ac{TE}.
The target embedding dimension $k$ can be optimised to account for cases where multiple past values are causal to the target.
An embedding delay $\tau$ can be used to better empirically capture the state from  a finite sample size \cite{Lizier2014}:
\begin{equation}\label{embedding_delay}
\mathbf{x}_{i}^k = \left\{x_{i-(k-1)\tau}, \ldots, x_{i-\tau}, x_{i}\right\}.
\end{equation}

The \ac{JIDT} package provides an option to perform an automated search for the optimal embedding dimensions and delays for the source and target variables when the \ac{KSG} estimator is used.
The use of auto-embedding was evaluated on test cases and significantly reduces the absolute magnitude of results obtained.
Sub-harmonic peaks were also removed and the optimising delay was preceded by a plateau instead of forming a well-defined peak.
If any auto-embedding parameter resolves at an edge of the allowed search space it should be widened.

\subsection{Parameter selection}\label{parameter_selection}

\subsubsection{Sample size}
In this context sample size refers to the number of data points involved in a single bivariate partial \ac{TE} estimation.
The main focus in selecting this parameter should be on identifying the smallest number that will allow for accurate and stable estimation of the conditional \ac{PDF}s involved.
Computation time scales exponentially with sample size depending on the number of embedding dimensions involved.

Small sample size leads to increased variability and varying absolute values in the case of box-kernel and \ac{KSG} estimators respectively.
The complexity and stability of process signals will dictate the minimum number of samples needed in order to estimate a representative \ac{PDF}.
Highly non-linear or directional (large condition number as obtained by \ac{SVD} analysis \cite{Skogestad2005}) processes will result in a more complex relationship between the signals and therefore a more complicated conditional \ac{PDF} in need of approximation.
In exceptionally difficult cases the practitioner might need to evaluate estimator convergence with regards to varying sample size to make an appropriate selection of this parameter.
In test cases estimated values converge for sample sizes above 2\,000.
This recommendation can also be found in prior art \cite{Bauer2005a}.
It might be worthwhile to investigate the effect of sample size on the specific process studied and compromise accordingly.

The desire to analyse an extensive period to capture an effect that occurred at an uncertain time should never motivate sample size selection.
Instead, performing \ac{MTR} analysis is recommended as this will be significantly faster, scaling linearly with total samples covered, and will also more accurately pinpoint the exact time at which a disturbance or fault entered the system \cite{Zhao2012}.

\subsubsection{Sampling period}
The product of the sample size and sampling period define the time span covered by a single estimation.
It is highly unlikely that time series data captured at a resolution faster than one second is available in a plant historian and the suggested sample size of 2\,000  points translates into a minimum time span of just over half an hour covered.
The sampling period is the primary parameter that should be varied to adjust the time span as sample size will always be constrained to a few thousand data points due to its impact on computational complexity.
However, the selected period should always satisfy the Nyquist sampling theorem for the frequency region of interest \cite{Skogestad2005}.
However, as the time constants of some processes, such as distillation columns, can be several hours, it might be useful to perform analyses on data sets sampled at lower frequencies.

A \ac{FFT} analysis can be used to assist with identifying time constants of interest.
If more than one frequency region is active, it is likely that this is due to the presence of two distinct disturbances.
Band-gap filtering the time series data and analysing each frequency region in isolation is recommended.
Disturbances with faster dynamics are usually more troublesome than slower ones.
However, as slow disturbances that go unchecked can result in oscillations of growing magnitude leading to the plant becoming unmanageable after a few days, it might be worthwhile to schedule analyses targeting slow dynamics periodically.

\subsubsection{Delay optimisation range}
Ideally, a large range of delays should be analysed at high resolution, but due to processing time constraints, compromises need to be made.
The dead-time of processes can vary significantly.
The throughput of a typical petrochemical plant is rather high compared to the hold-up volumes, and therefore the maximum time delays are expected to be relatively small.
However, this is not necessarily the case for mineral and pharmaceutical plants.
The initial delay optimisation search space covers should cover at least five minutes.
If many directed variable pair weights maximise at the search space borders, the analyst should increase the delay range.

The optimising time delay is a property of the processes involved and not likely to change significantly under the influence of different faults and disturbances.
Over time, the ideal delay range that optimises each particular directed variable pair for a specific system will become known.
The computational load can be reduced by detecting an appropriate time delay range for each directed variable pair and narrowing the search space accordingly.
Using mutual information instead of \ac{TE} to determine the optimal delay range is another successful strategy for reducing computational load \cite{Su2017a}.

\subsection{Significance testing}

\ac{TE} estimation methods are likely to return non-zero values for any set of random signals.
As the proposed scheme for fault and disturbance screening is fully data-driven, unsupervised and model-free, no first principle or model-based knowledge can be used to force any meaningful structure to the results obtained.
Significance testing is necessary to confirm whether an estimated \ac{TE} value is due to real interaction.

\subsubsection{Magnitude}
A significance threshold on the magnitude of a bivariate \ac{TE} estimate is typically derived by studying the distribution of \ac{TE} estimates obtained on a surrogate set of data generated such that the temporal ordering between the original data vectors is broken while maintaining as much of the other properties of the signal as possible.
Some researchers simply random shuffled the source data vector in time \cite{Marschinski2002, Lizier2011} while others advocated the use of the \ac{iAAFT} method \cite{Schreiber2000a} that strives to retain amplitude both the amplitude distribution and power spectrum \cite{Bauer2005a}.

Numerous researchers applying \ac{TE} to process data derive a significance threshold based on a certain number of standard deviations away from the mean of values calculated on the surrogate set \cite{Bauer2005a, Su2017a}.
However, this practice is discouraged for non-linear metrics and non-parametric rank-order approaches are recommended instead \cite{Schreiber2000a}.
According to a rank-order test, a residual probability of a false rejection $\alpha$ corresponds to a significance level of $\left(1 - \alpha \right) \times 100 \%$.
A test for \ac{TE} significance in a specific direction of time is one-sided for which $M =  1/\alpha - 1$ surrogate sequences are generated, giving a total of $1/\alpha$ data sets including the data set itself.
The probability that the original data results in the largest causality measure value by coincidence is, therefore, $\alpha$.
For a significance level of 95\%, the test thus generates and calculates the \ac{TE} of 19 surrogate data sets and requires that the real data returns results higher than all the surrogates to pass.

\subsubsection{Directionality}\label{directionality_test}
Although \ac{TE} is an asymmetric measure and therefore implicitly tests significance in a specific direction, automated optimisation over a range of delays complicates matters as an affected variable might appear to influence their causal variables if the affected variable value is lagged far enough behind the causal variables.

An optional additional directionality significance test is applied to help reduce spurious connections detected due to this effect.
The test involves calculating the \ac{TE} over a range of delays shifting the source target forward and backward, recording the maximum values and their associated delays in each direction: $\left[\psi^{\text{forward}}; \delta^{\text{forward}}\right]$ and $\left[\psi^{\text{backward}}; \delta^{\text{backward}}\right]$ respectively.
The directionality test passes on condition that $\psi^{\text{backward}} > \psi^{\text{forward}}$ or $\delta^{\text{backward}} < \delta^{\text{forward}}$.
Note that application of this test will suppress the detection of true bi-directional interactions.

\subsection{Scaling}
Ideally the value of entropy measured on any signal should have a consistent practical interpretation across variables regardless of the engineering units used to record the measurements.
However, the estimated entropy of a continuous signal using kernel methods is dependent on the relationship between its variance and the kernel bandwidth used to estimate the underlying \ac{PDF}.
To ensure that the magnitude of estimated \ac{TE} between variable pairs has a consistent meaning throughout an analysis, the scaling of data before estimation of \ac{TE} is critical when using kernel estimators.


Even though standardisation (subtraction of the mean and division by the standard deviation) is a commonly applied scaling strategy \cite{Bauer2005a, CastanoArranz2012} its use removes all information about how much the signal varies relative to its desired operating range and thereby introduces the requirement that sufficient and representative excitation of the process in all directions should be present in the sample data.
If this is not the case, division by the small standard deviation will blow a small amount of noise on an otherwise quiet signal out of proportion, misrepresenting the relative real information content carried by the signal.
For example, we do not wish to flag a disturbance that causes a level (an unimportant inventory variable) to fluctuate between 48 and 52\% while the desired control limits are 30-80\% as critically important.
However, a fault that causes causing an analyser composition reading (an economic variable) to fluctuate between 0.85 and 0.86 while the minimum requirement is 0.84 should be regarded as highly significant.

Strong cases for the necessity of getting a scaling independent representation of data to make reliable conclusions about relative interaction strengths among process elements using weighted graphs have been made elsewhere \cite{CastanoArranz2012}.
A suitable scaling strategy involves subtracting the nominal operating value and division with the maximum expected or allowed change \cite{Skogestad2005, CastanoArranz2012}.
If the nominal operational point does not lie in the middle of the allowed or desired range, the most conservative approach is to use the biggest difference to scale disturbances and reference (set-point) values, while the smallest difference is used to scale allowed input changes and control errors.
The result is that all values have less than unity magnitude if they remain within the predetermined limits informed by process knowledge, and similar magnitudes indicate a deviation of equal severity.

Both standardisation and process limit base scaling was evaluated on test cases.
The \ac{KSG} estimator has an adaptive bandwidth and is not sensitive to the scaling approach used while a significant improvement in ranking results is observed when using the kernel estimator which depends on a fixed bandwidth parameter.

\section{Ranking nodes using graph centrality}

\subsection{Eigenvector centrality}

Eigenvector centrality measures a node's influence in a weighted directed graph as the sum of importance scores nodes with an edge incident to it multiplied by the edge weights,

\begin{equation}\label{eq:sum_of_importances}
C_{\text{eig}}(k) = \sum_{j\in\mathbf{L}_k} w_{kj} x_j,
\end{equation}
where $C_{\text{eig}}(k)$ is the importance of node $k$, $\mathbf{L}_k$ is the set of nodes with connections to $x_k$ and $w_{kj}$ are entries of the edge weight matrix $\mathbf{W}$.
In the context of this application the edge weight matrix $\mathbf{W}$ must be column-stochastic (all columns should sum to one) with real positive entries representing a measure of connection strength between nodes.
The problem can be rewritten in the format of a standard (right) eigenvalue problem

\begin{equation}\label{eq:eigenvector_solution}
\mathbf{Wx} = \lambda\mathbf{x}.
\end{equation}

Although there might be many eigenvalues $\lambda$ with different eigenvectors $\mathbf{x}$ which satisfy this equation, the eigenvector with all positive entries and associated with an eigenvalue of unity, $\lambda = 1$, contains the relative importance scores.
This eigenvector corresponds to the stationary probability vector defined on a stochastic matrix and the Perron-Frobenius theorem guarantees its existence and uniqueness.

\subsection{Markov chain interpretation}

The eigenvalue centrality measure can be interpreted as the results of a Markov chain operator on the graph that assigns node importance relative to the distribution of time a random walk starting on any node will spend on the various nodes.
The proof relates to the idea that the long-term probability of being in a state is independent of the initial state.
A Markov chain problem formulation usually consists of a \ac{TPM}  $\mathbf{G}$ (also called a substitution matrix) together with a \ac{RPM} $\mathbf{Q}$ weighed by a mixing factor $m$
\begin{equation}\label{eq:weighted_matrix}
\mathbf{M} = m \mathbf{G} + (1-m) \mathbf{Q}.
\end{equation}

Entries in the \ac{TPM} indicate the probability that an actor currently residing on a specific node will move to a neighbouring node -- in the appropriate direction for the case of directed graphs.
The \ac{RPM} represents the probability that an actor will move randomly to any node on the network.
In basic algorithms, this probability distribution is usually considered to be equal among all nodes.
Since the sum of all probabilities for a given event should be unity, both the \ac{TPM} and \ac{RPM} have stochastic column vectors containing non-negative real numbers.

For the Markov chain operator interpretation to hold, it is essential that there be no dangling nodes, that is, columns in $\mathbf{M}$ which have no non-zero entries (whose sum is, therefore, zero).
In the context of this work dangling nodes are those that do not influence any other, a situation that is very likely to occur.
A straightforward means of dealing with dangling nodes is to add very weak connectivity between all nodes in the network.
Weighing a \ac{TPM} $\mathbf{G}$ with an equal probability \ac{RPM} $\mathbf{Q}$ with a $m$ very close to unity implements this solution.

\subsection{Direction of analysis}

There are two different directions in which the eigenvector centrality can be used to assess importance.
Some applications, such as the ranking of web pages, attribute importance based on the quantity and significance of \emph{inbound} connections to a specific node, or references back to a particular page.
In the context of determining which node has the most considerable extent of causal influence on other nodes, it is desired to attribute importance based on the quantity and significance of \emph{outgoing} connections from a specific node.

The convention used to define entries in the adjacency or weight matrices according to row or columns and whether the right or left eigenvector problem is solved determines if the ranking results refer to the inbound or outbound metric.
In this work, eigenvector centrality always refers to the results obtained by calculating the right eigenvector with the adjacency matrix arranged source (causal) nodes along the columns and destination (affected) nodes along the rows.
Under this arrangement it is necessary to use the transposed weighted digraph adjacency matrix $\mathbf{M}$ as the ranking weight matrix $\mathbf{W}$ in \ref{eq:eigenvector_solution} to get the desired result of ranking based on outgoing connections
\begin{equation}\label{eq:transposed_weightmatrix}
\mathbf{W} = \mathbf{M^\intercal}.
\end{equation}

\subsection{Introducing bias}\label{ranking_bias}
Identification of the most critical faults might be enhanced by incorporating individual controller performance \ac{KPI}s and stream importance measures, such as economic value, as meta-data in the node ranking problem.
Adjusting the relative reset probabilities is an ideal means of biasing the network ranking towards specific nodes.

A relative reset bias vector $\mathbf{q}$ needs to be provided and normalised such that it sums to unity.
The relative, not absolute, values of the entries is of importance.
The reset probability matrix $\mathbf{Q}$ is then defined as the $n \times n$ matrix with each row equal to the normalised reset bias vector $\mathbf{q}$:

\begin{equation}
\mathbf{Q} = \left[ \begin{array}{lll}
\mathbf{q}, \mathbf{q}, \cdots \end{array} \right]^\intercal.
\end{equation}

The transition and reset probability mixing weight $m$ can be considered a tuning factor that will determine how significant the edge weightings are relative to the node biases in determining the final ranking score.

\subsection{Weight normalisation}
Edge weights should be normalised to maintain a consistent interpretation of the ranking results independent of the units used to record connection strength and to provide a scale with regards to the (biased) reset probability weights.
Additionally, for $\mathbf{W}$ to be column stochastic the rows of $\mathbf{M}$ must sum to unity.
To preserve the meaning of reset probability, $\mathbf{G}$ needs to be row-normalised before linearly combining with $\mathbf{Q}$ to form $\mathbf{M}$.
Row normalisation involves dividing each element of a matrix by the sum of the elements in the same row
\begin{equation}\label{eq:row_norm}
X_{ij_{\text{norm}}} = {\left(\frac{X_{ij}}{\sum \limits_{i} X_{ij}}\right)}.
\end{equation}

If dangling nodes are present in the network, some rows of $\mathbf{G}$ will sum to zero instead of unity.
These rows will not result in a row-normalised weight matrix $\mathbf{M}$ after linear combination with $\mathbf{Q}$ and a final row-normalisation step is needed to ensure that $\mathbf{W}$ is column-stochastic.
unities for future research, see Section~\ref{fut:metadata}.

\subsection{Katz centrality}\label{katz_centrality}
Katz centrality measures the total number of walks (sequence of edges and nodes connecting two nodes) between two nodes in the network.
In effect, it is similar to eigenvector centrality with the additional accounting for higher order connections weighed by an attenuation factor and can assist with boosting the difference in ranking scores between nodes.

The formal definition of Katz centrality is
\begin{equation}\label{eq:katz}
C_{\text{Katz}}(k) = \sum_{p=1}^{\infty} \sum_{j=1}^{N} \alpha^{p} (W^{p})_{kj},
\end{equation}
where $C_{\text{Katz}}(k)$ is the importance of node $k$ according to the Katz centrality measure, $\mathbf{W}$ is the weight matrix and $\alpha$ is as an attenuation factor that cannot exceed a value of $1/\lambda$ where $\lambda$ is the largest eigenvalue of $\mathbf{W}$.
Unlike eigenvector centrality, it is possible to compute Katz centrality on directed acyclic graphs (DAGs), which removes the need for adjusting the adjacency matrix with a fully connected reset matrix.

The principal eigenvector associated with the largest eigenvalue of  $\mathbf{W}$ is returned in the limit that $\alpha$ approaches $1/\lambda$ from below.
This is identical to the result calculated according to eigenvector centrality when an acyclic graph is combined with a uniform reset bias vector with $m$ approaching zero from the right.
The inclusion of weak indirect connections in \ac{ITN}s by partial bivariate \ac{TE} estimators is comparable to calculating a ranking based on Katz centrality.

\section{Recommended procedures}
Efficient procedures for applying the presented method will depend on the fault or disturbance's nature and how much the analyst already knows about it.
Applications can range from an original assessment of the key influences in a plant to confirmation of already existing hypotheses concerning propagation pathways.
The intended use of the presented method is as a triaging aid for use by process experts and is not refined enough for drawing blind conclusions, but should be integrated into a larger \ac{FDD} workflow.

It is difficult to provide performance guarantees under a range of different conditions for unsupervised methods.
Instead, sanity checks are required to eliminate spurious results, and visual graph interpretation is occasionally necessary to locate the actual source of information flow.

\subsection{Variable selection}
Not all data sources associated with a specific section of a plant will contribute useful information to the analysis, and it is recommended that they are filtered to reduce the scope to no more than 50 elements.
If it is desired to analyse a specific disturbance event or fault, care should be taken to ensure that it is included in the period covered by the actual calculations.
For complex processes a high-level analysis of the most important economic variables can be used to inform more detailed, focused analysis of flagged process sections.

Unchanged set-points and other constant data vectors will not play a role in any causal relationship, and the analyst can safely eliminate these from the data set without any loss in fidelity.
The analyst may also eliminate redundant measurements upon confirming that no sensor faults took place during the period under investigation.
Measurements at the very end of the process stream not connected via any controllers or other feedback mechanisms only have to be included if they are affected.

\subsection{Multiple time region analysis}
A single snapshot of network ranking scores might be difficult to interpret without comparison to nominal fault-free operation.
Monitoring the relative changes in node rankings over time might provide more insight.
A plot of node importance calculated for multiple highly overlapping rolling windows of time provides a single comprehensive overview of changes in the process' causal structure and is considered the primary output of the proposed method for the purpose of incipient fault detection.

\subsection{Visual inspection}
Visual inspection of the \ac{ITN} while continually referring to a \ac{PFD} or basic description of the process, such as is commonly found in the operator training manual, is arguably the most useful means of interpreting the results and forming hypotheses about the source of faults and disturbances.

Colour grading nodes based on their centrality scores make it easy to visually process their relative locations and the areas they influence.
However, if the indicated nodes are unexpected or no clear link between them and known faults is apparent this information is of limited use in isolation.
If the most important edges are highlighted it becomes significantly easier to understand why some nodes have high centrality scores.
The associated software saves the \ac{ITN} in \verb+GML+ format.
Cytoscape \cite{Shannon2003} is an open source and user-friendly tool for viewing and analysing large networks and supports applying the analysis steps mentioned above.

An appropriate layout enables visual interpretation of a network's structure.
In the context of tracing an effect to its source it is desirable to arrange nodes from most independent to dependent.
A hierarchical network layout serves this purpose.
Hierarchical layouts separate the nodes into clusters influenced by different original information source nodes, making the method applicable to situations that involve multiple distinct faults.
The yFiles hierarchic network layout algorithm provides the most useful results amongst the options available in Cytoscape.

A hierarchical layout cannot be computed on a fully connected graph as produced if the recommended eigenvector centrality method is applied.
Deleting all edge weights below a certain percentile or threshold is usually necessary for visualisation purposes.

\section{Results}

\subsection{Simple mixing process}

\begin{figure}[htbp]
	\centering
	\includegraphics[width=0.5\textwidth]{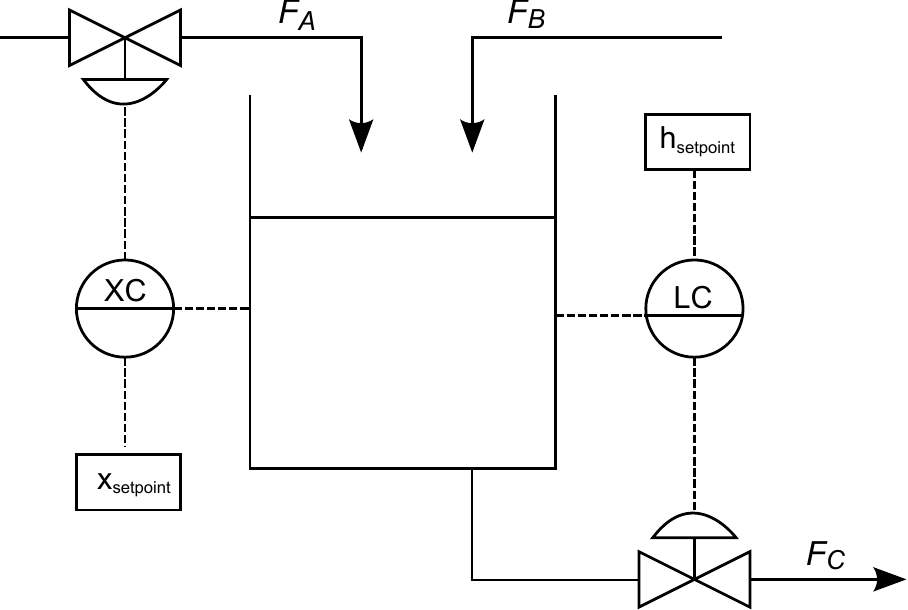}
	\caption{Level and composition problem process flow diagram}
	\label{fig:lc_schematic}
\end{figure}

\begin{figure}[htbp]
	\centering
	
	\begin{tabular}{cc}
		\subfloat[standardised]{
			\includegraphics[width = 0.5\linewidth]{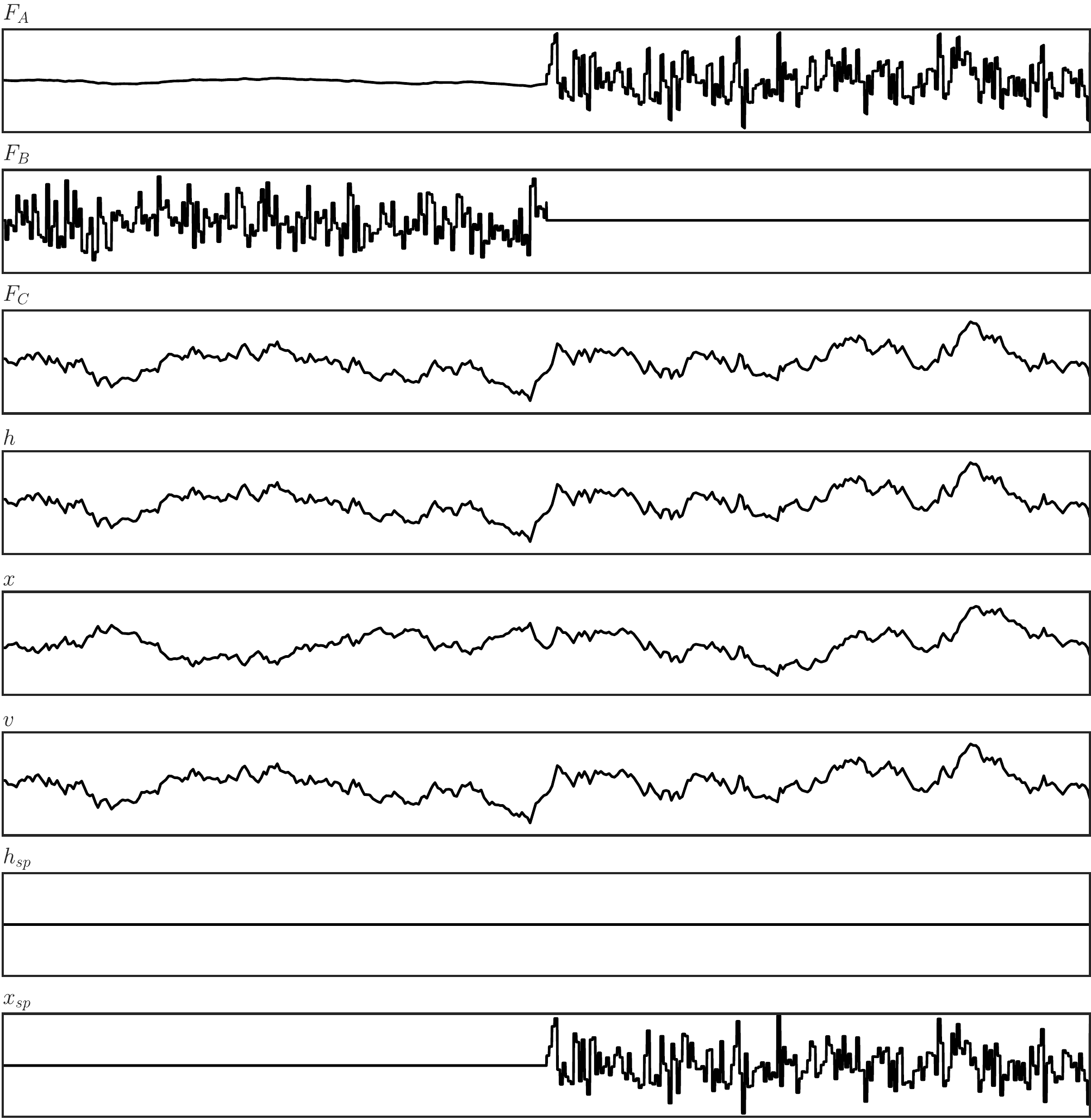}} &
		\subfloat[process limit scaled]{
			\includegraphics[width = 0.5\linewidth]{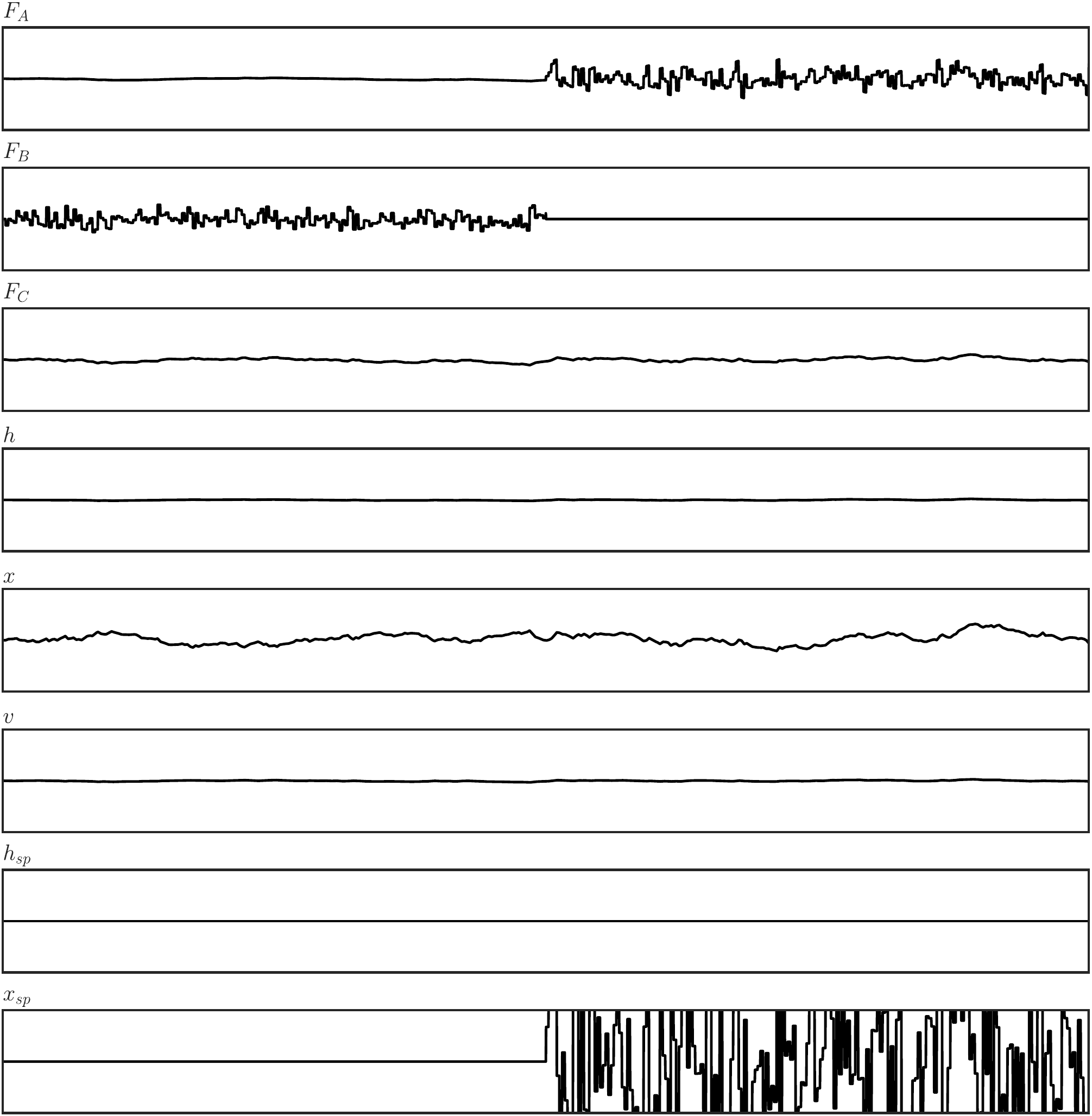}} \\
	\end{tabular}
	
	\caption{High density time series plot of disturbances}
	\label{fig:level_concentration_FB_xsp_dist_hdts}
\end{figure}

The simple mixing process presented in Figure~\ref{fig:lc_schematic} exhibits many of the aspects we wish to study, including non-linearity, interacting control and integrating processes.
To assess the efficiency of different methods, random noise is introduced in $F_B$ from $t=0$ until $t=20$, and the composition set-point is perturbed thereafter until the end of a 40-hour simulation.
Figure~\ref{fig:level_concentration_FB_xsp_dist_hdts} provides high density time series plots of the standardised and process limit scaled simulation results respectively.
Note the attenuation of outlet flow rate and height signals together with amplification of the composition signal in the latter.

\begin{figure}[htbp]
	\centering
	
	\begin{tabular}{cc}
		\subfloat[kernel estimator; standardised]{
			\includegraphics[width = 0.5\linewidth]{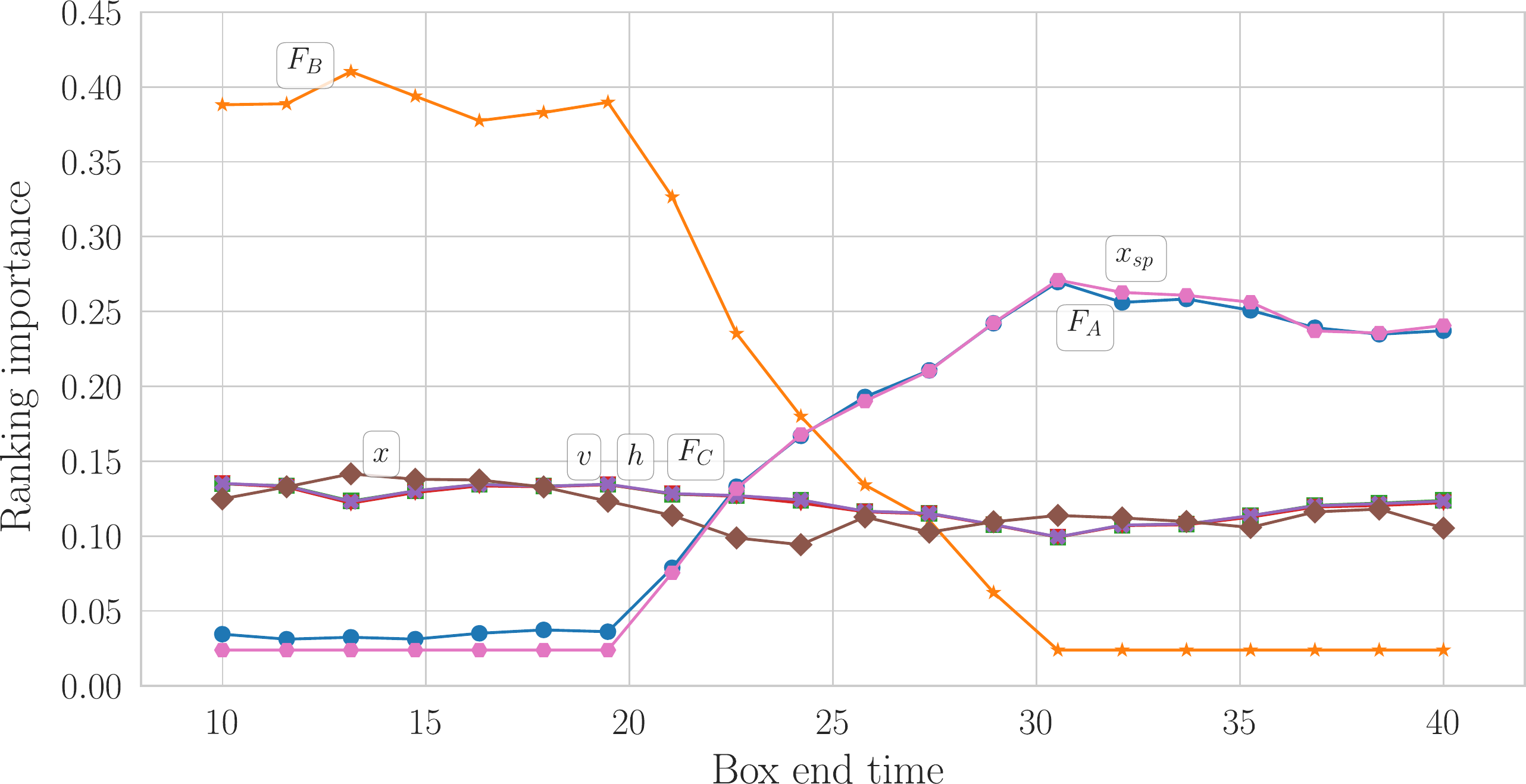}} &
		\subfloat[kernel estimator; process limit scaled]{
			\includegraphics[width = 0.5\linewidth]{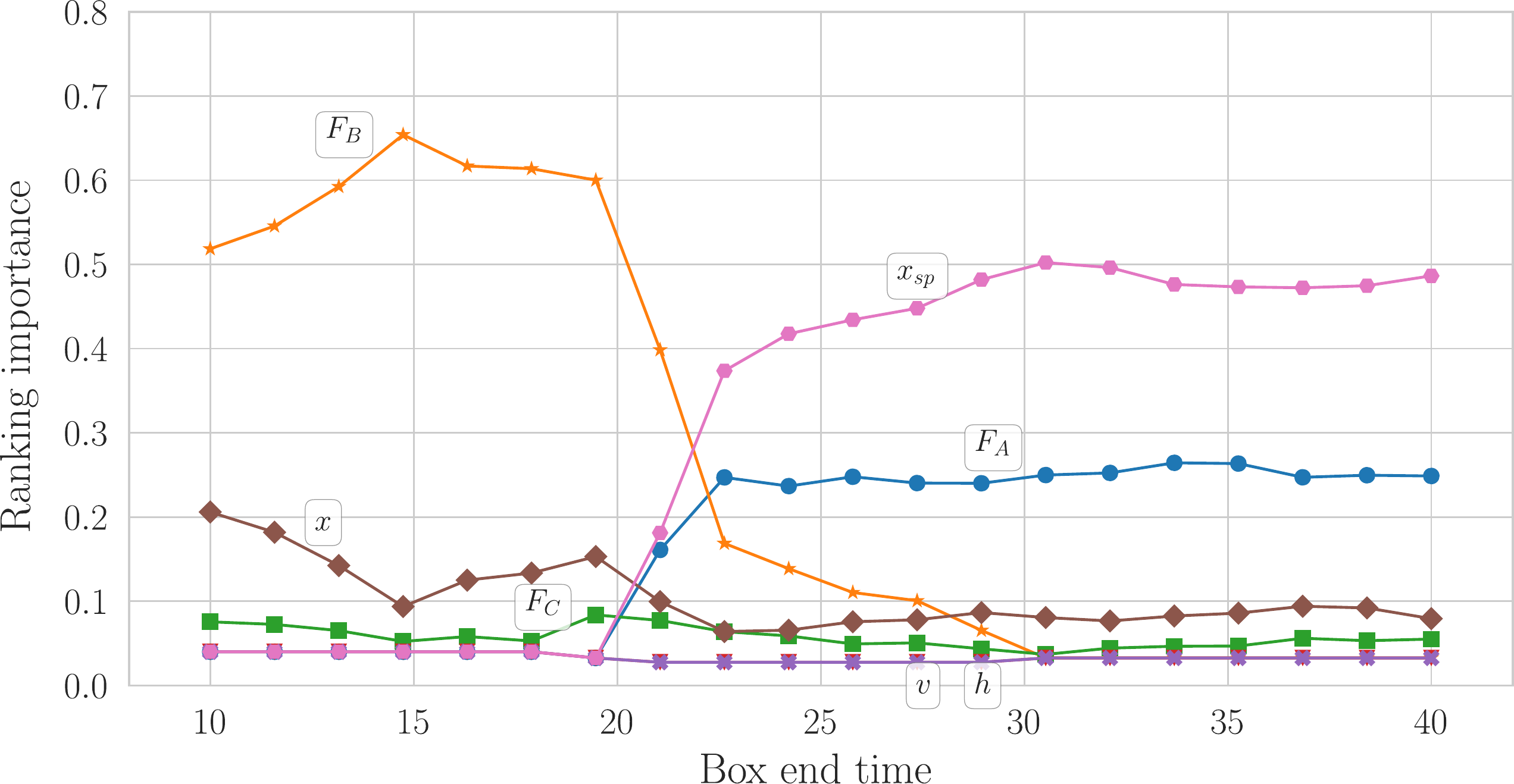}} \\
		\subfloat[KSG estimator; standardised]{
			\includegraphics[width = 0.5\linewidth]{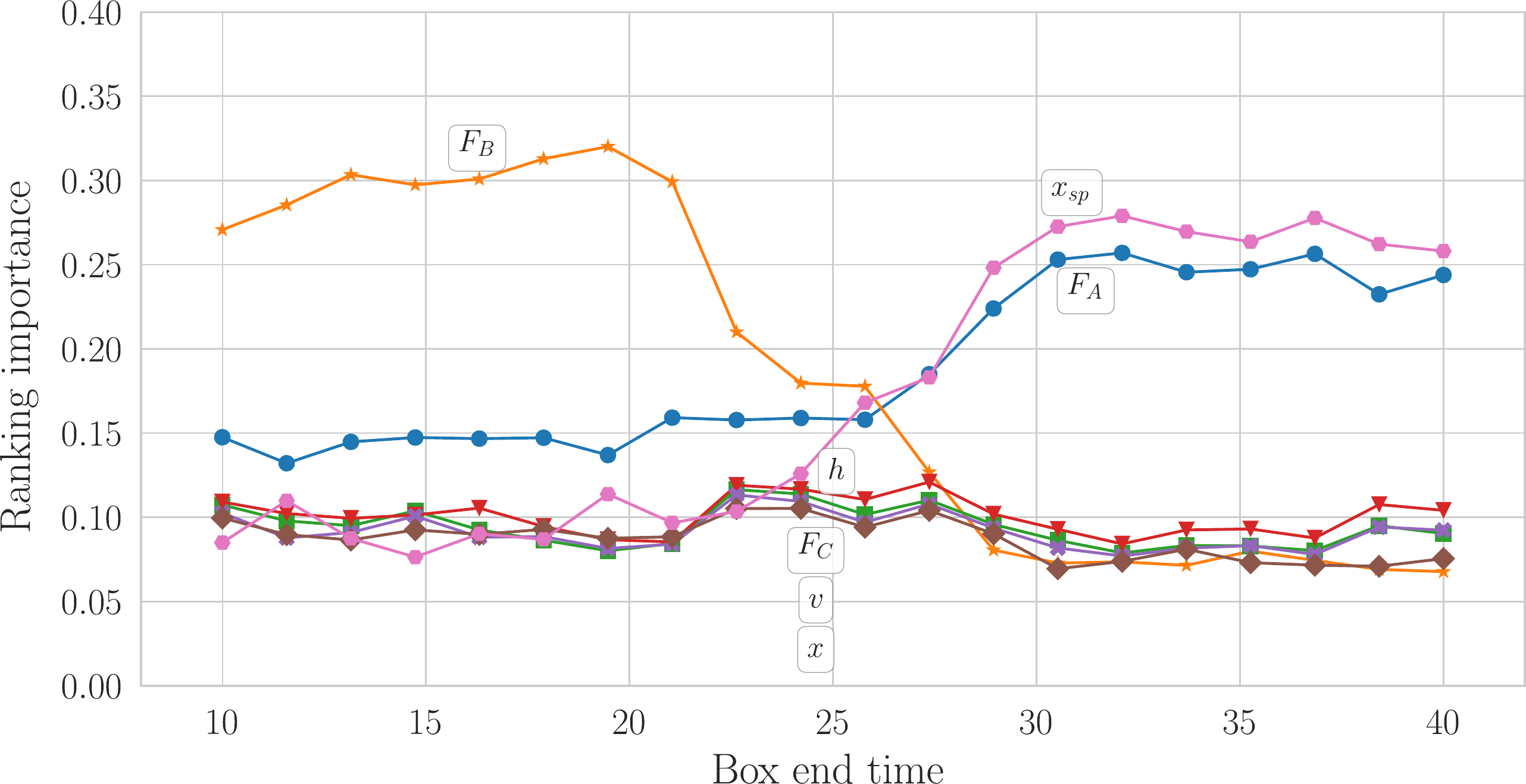}} &
		\subfloat[KSG estimator; process limit scaled]{
			\includegraphics[width = 0.5\linewidth]{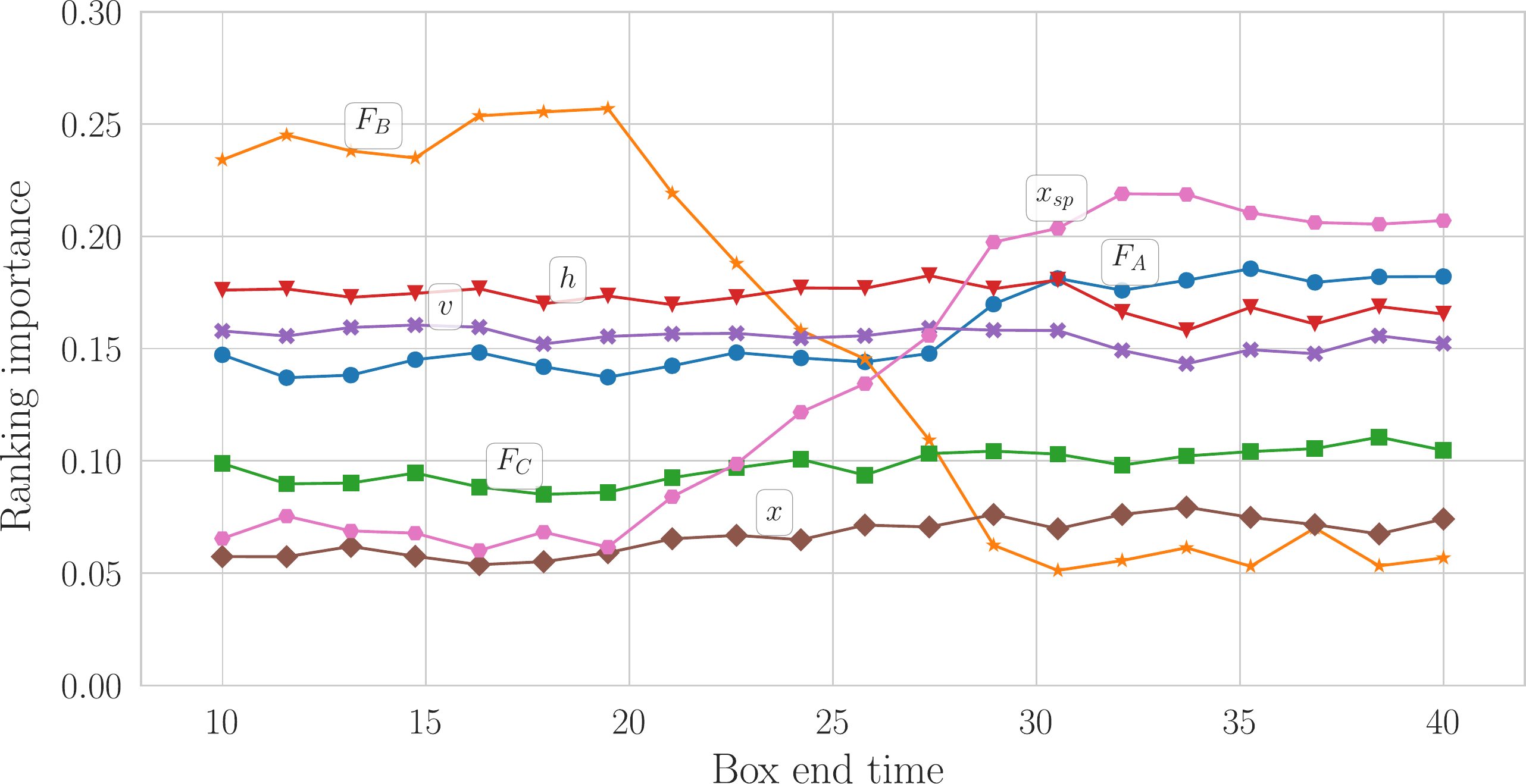}} \\
		\subfloat[KSG estimator; standardised; embedded]{
			\includegraphics[width = 0.5\linewidth]{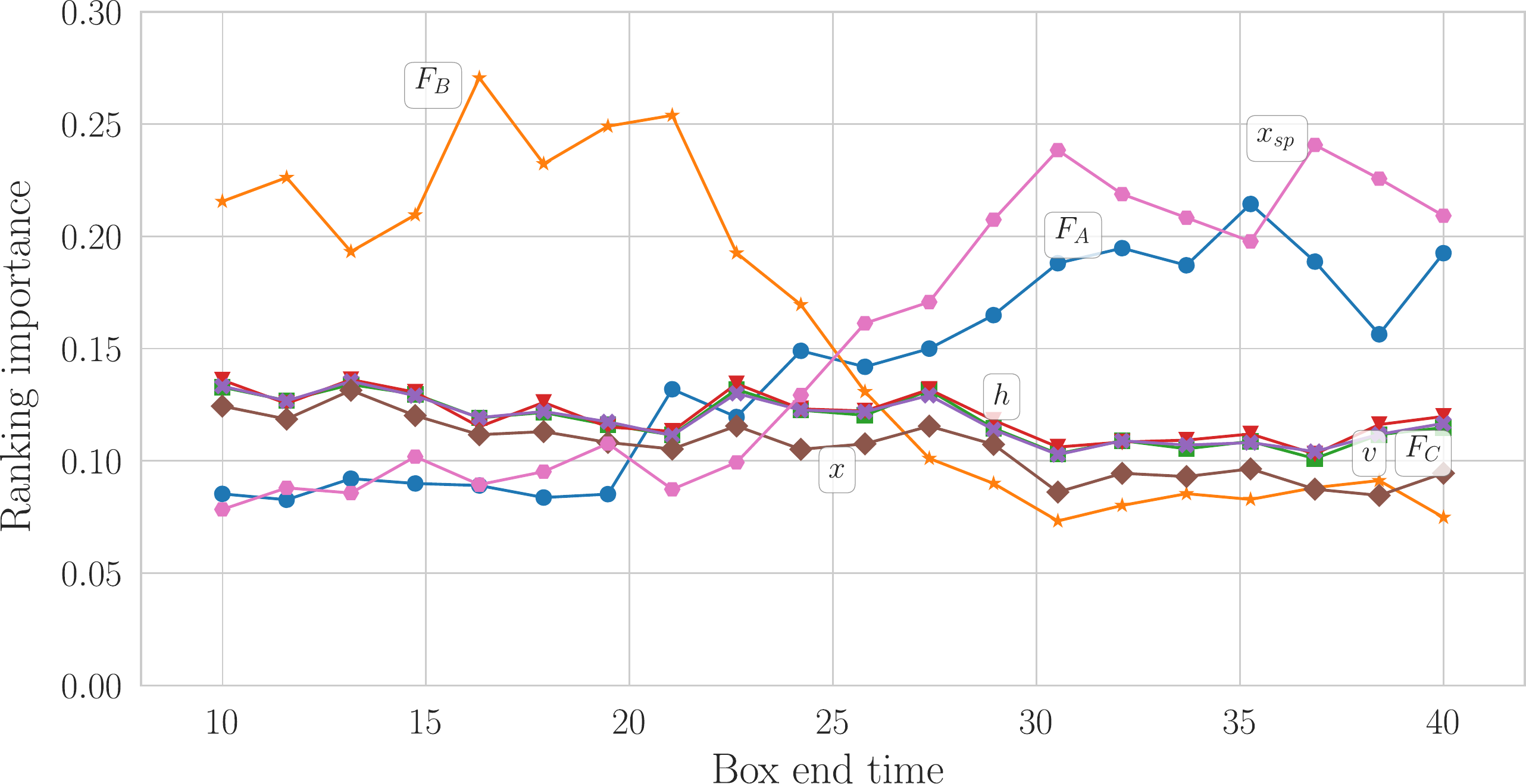}} &
		\subfloat[KSG estimator; process limit scaled; embedded]{
			\includegraphics[width = 0.5\linewidth]{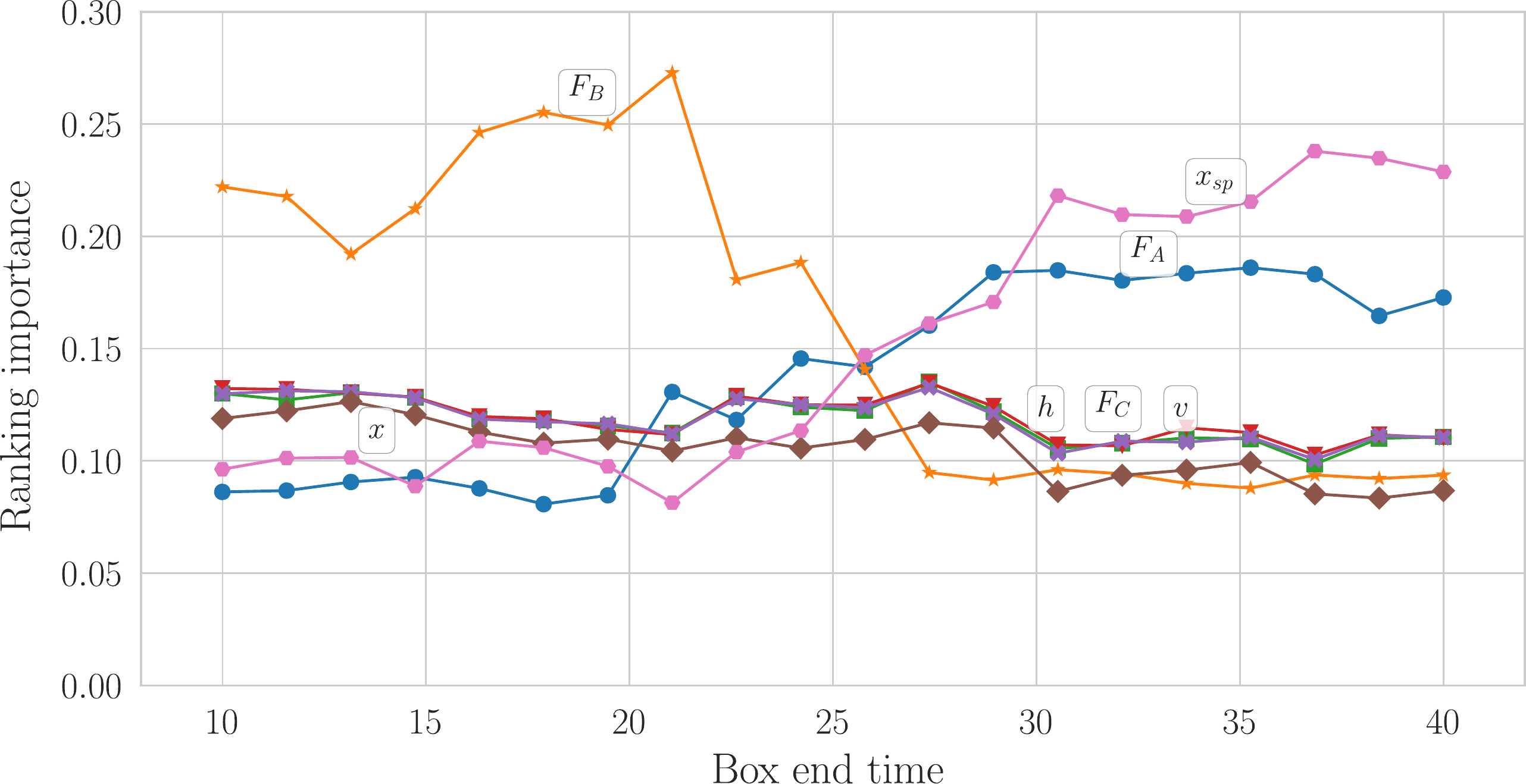}} \\
	\end{tabular}
	
	\caption{Multiple time region node rankings for mixing process [simple estimates; no significance testing]}
	\label{fig:level_concentration_FB_xsp_dist_nosigtest_simple}
\end{figure}

Figures~\ref{fig:level_concentration_FB_xsp_dist_nosigtest_simple} plots node importance scores over multiple time regions obtained without the use of significance testing.
Note how process limit scaling enables the kernel method to rank $x_{sp}$ higher than $F_{A}$, but made rankings less distinct when the \ac{KSG} method is used without embedding.
The use of directional \ac{TE} and significance testing did not result in better \ac{MTR} plots.

Figure~\ref{fig:level_concentration_xsp_dist_nosigtest_absolute_ITNs} shows \ac{ITN}s for the last time window investigated where the composition set-point disturbance has been active in isolation for a long time.
Significance testing was applied and the graphs have been further simplified by deleting the bottom half of edge weights and resolving higher-order connections up to the fifth degree of separation.
Note that the two constant signals of $F_{B}$ and $h$ passed a significance test in a number of cases involving the \ac{KSG} estimator.
This is due to the \ac{KSG} estimator adding a small amount of noise to signals in order to ensure differentiation between points, a requirement for its convergence.
We therefore conclude that it is important to remove stationary signals before analysis with the \ac{KSG} estimator.
Another obvious mistake is that although $x_{sp}$ is an independent source of information, some methods indicate incoming connections.
This is due to the delay optimisation lagging the real affected variable behind the source, and applying the bi-directional significance tests described in Section~\ref{directionality_test} might eliminate these errors.
In general, the \ac{ITN}s produced by the \ac{KSG} estimator are a better resemblance of reality, and process limit scaling allows for better detection of propagation paths resulting in a more hierarchical graph (compare Figure~\ref{fig:level_concentration_xsp_dist_nosigtest_absolute_ITNs} (g) and (h), for example). 

The significant contrast between the accuracy of individual \ac{ITN}s and the \ac{MTR} ranking results clearly demonstrate the significant benefit of performing node centrality analysis for the purpose of fault detection.
Ranking does not only condense information but also filters out the effect of individual erroneous connection by aggregating over all the results.
A similar result is seen in boosting models for regression and classification, where an aggregate of mediocre models manages to provide results that are better than any individual model.

\begin{figure}[htbp]
	\centering
	
	\begin{tabular}{cc}
		\subfloat[simple; kernel estimator; standardised]{
			\includegraphics[width=0.4\linewidth]{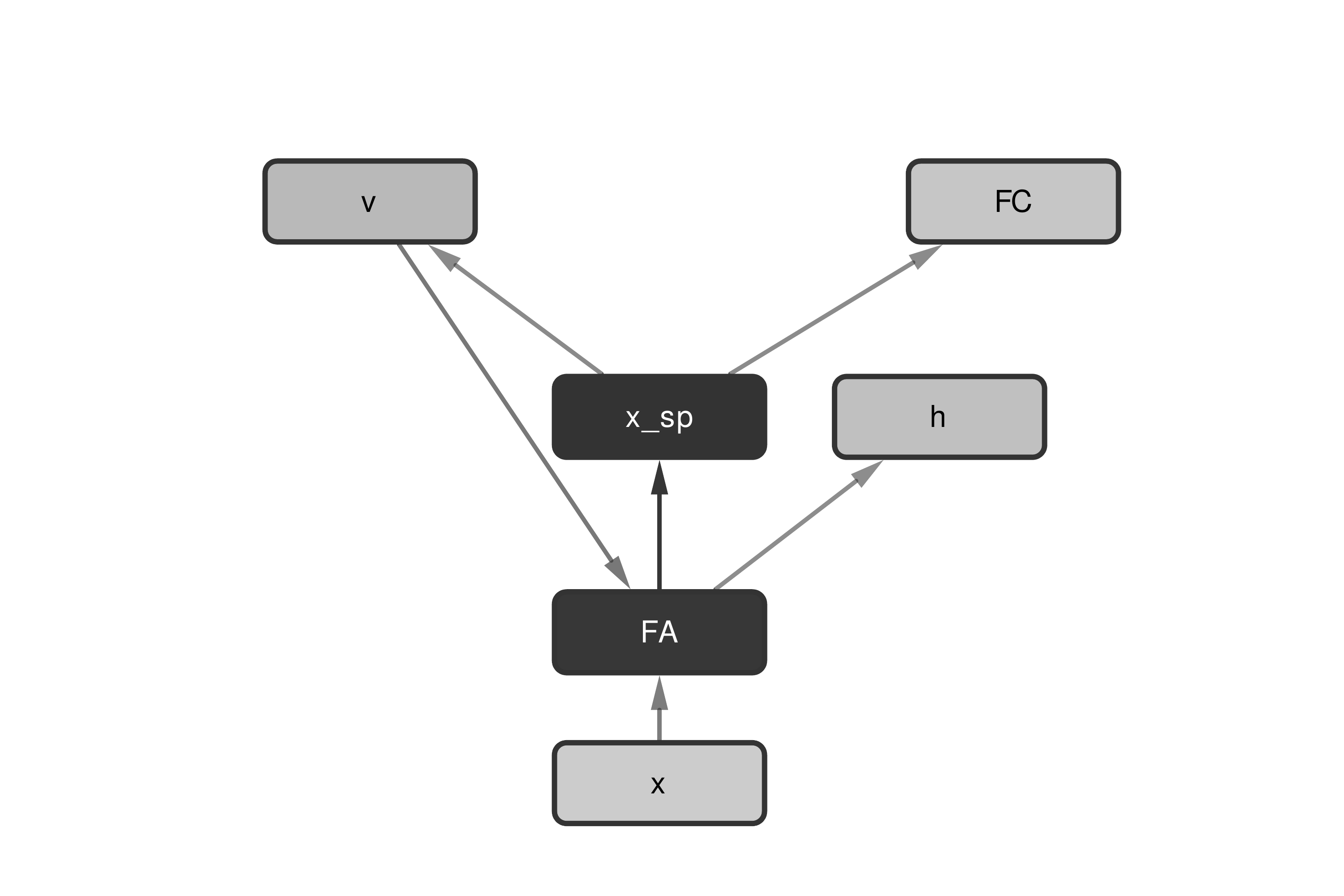}} &
		\subfloat[simple; kernel estimator; process limit scaled]{
			\includegraphics[width=0.4\linewidth]{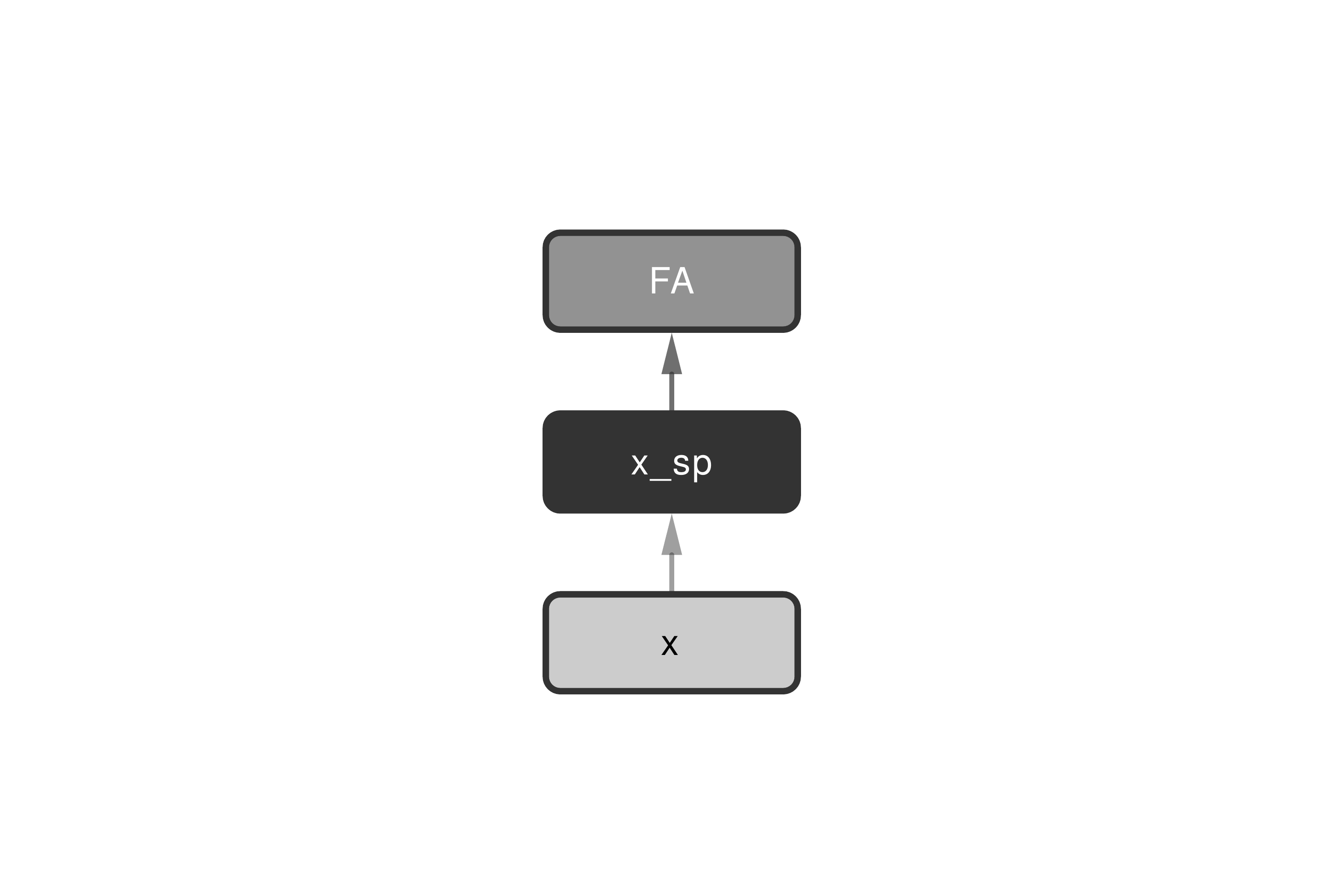}} \\
		\subfloat[simple; KSG estimator; process limit scaled]{
			\includegraphics[width=0.4\linewidth]{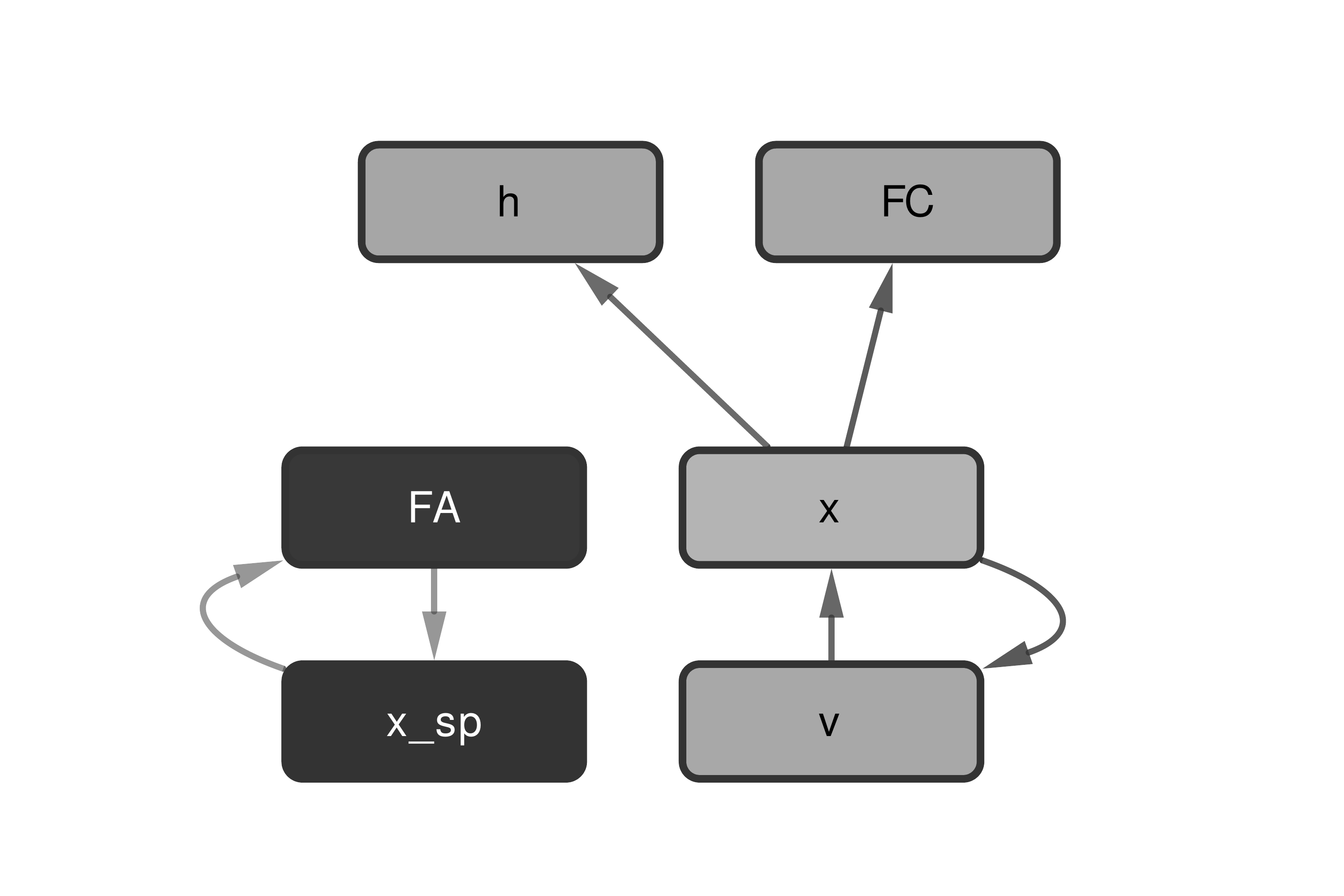}} &
		\subfloat[simple; KSG estimator; process limit scaled]{
			\includegraphics[width=0.4\linewidth]{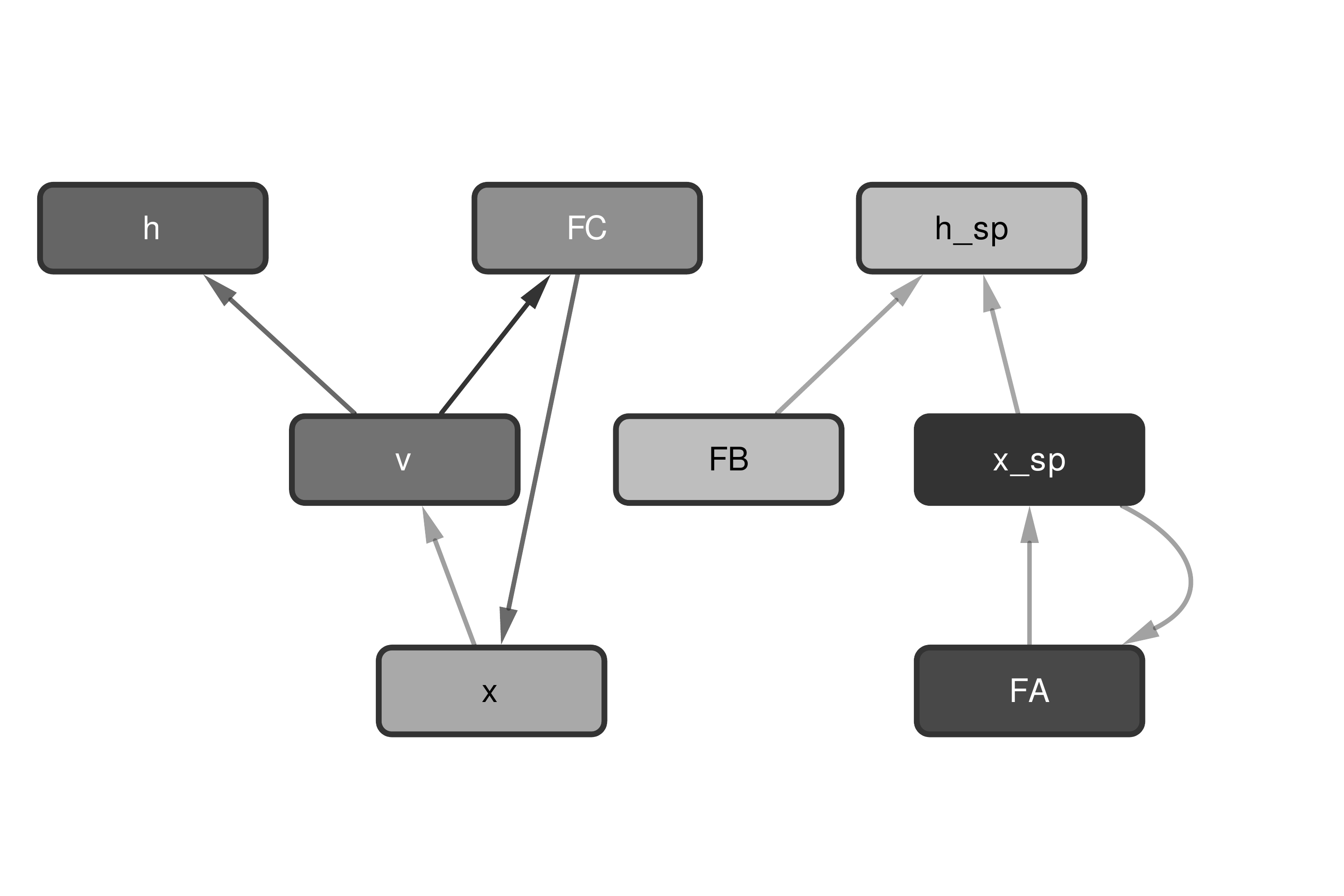}} \\
		\subfloat[directional; kernel estimator; standardised]{
			\includegraphics[width=0.4\linewidth]{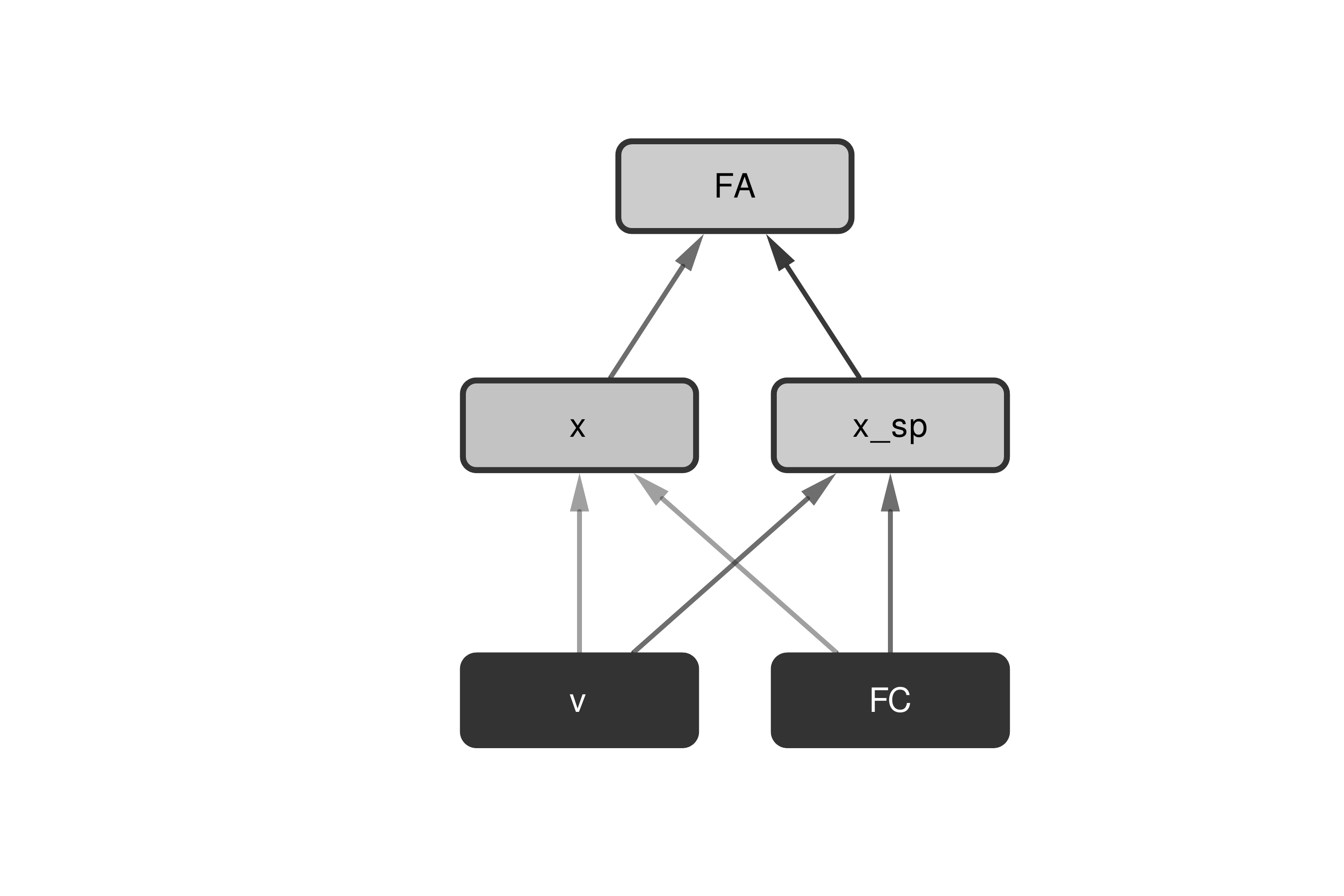}} &
		\subfloat[directional; kernel estimator; process limit scaled]{
			\includegraphics[width=0.4\linewidth]{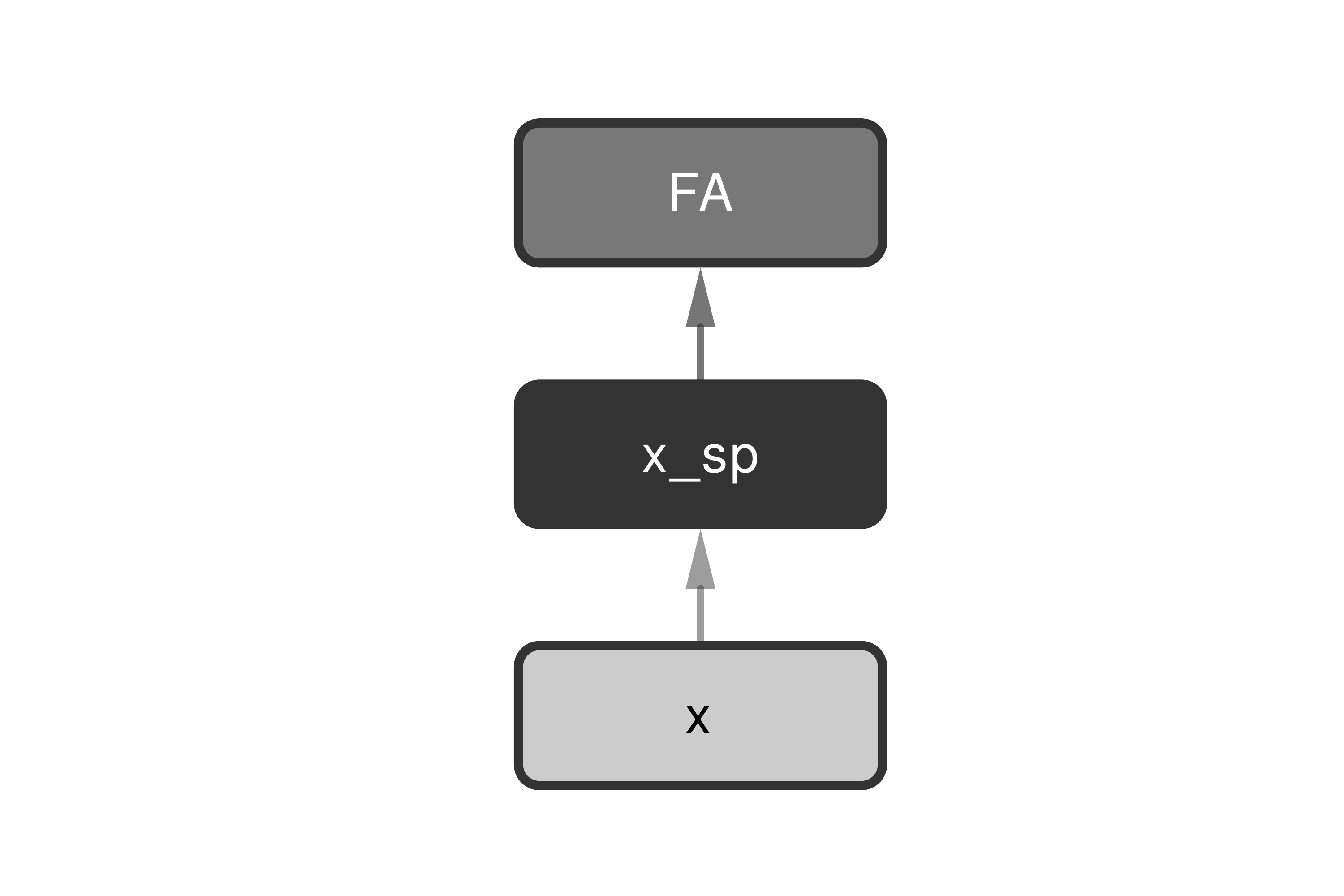}} \\
		\subfloat[directional; KSG estimator; standardised]{
			\includegraphics[width=0.4\linewidth]{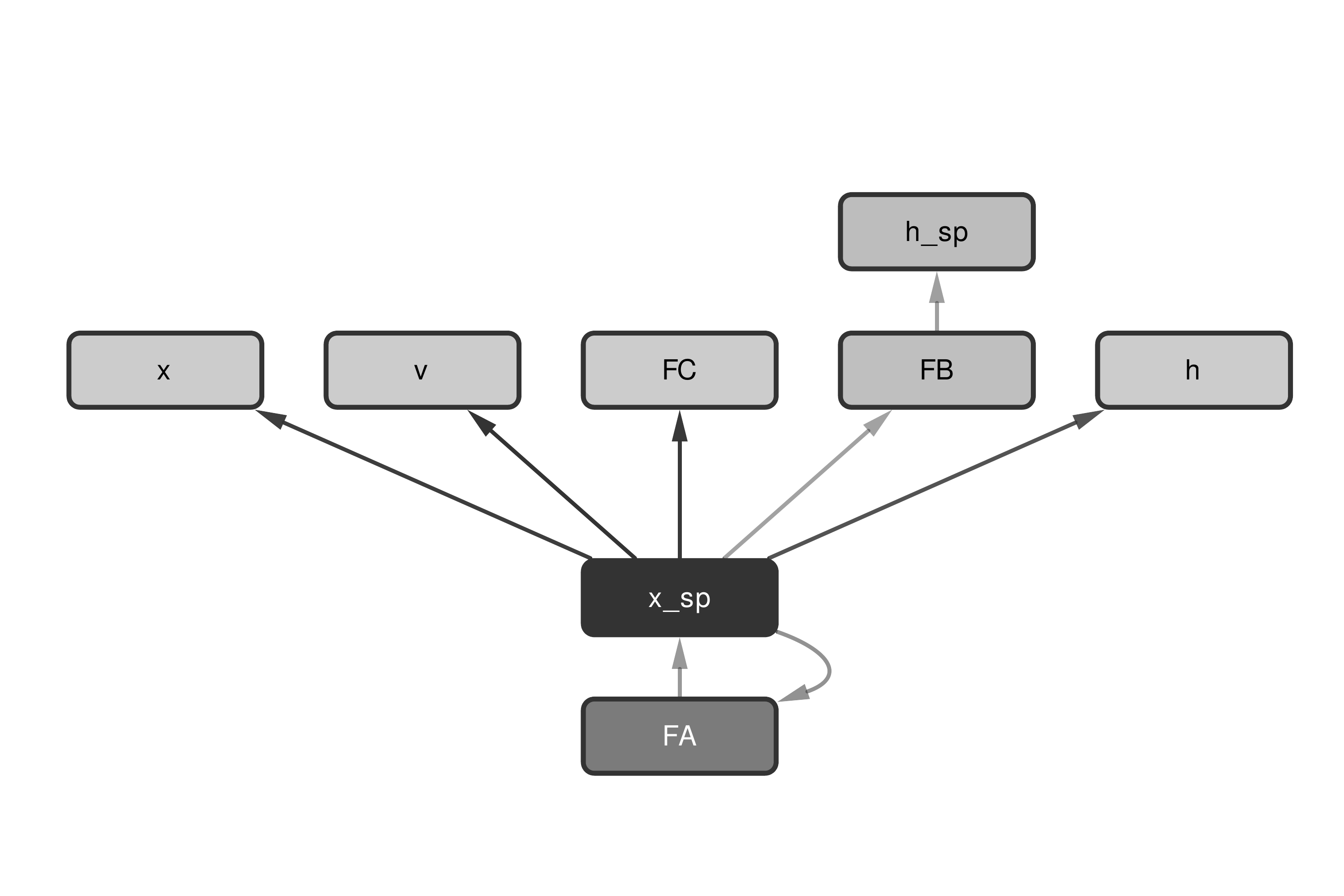}} &
		\subfloat[directional; KSG estimator; process limit scaled]{
			\includegraphics[width=0.4\linewidth]{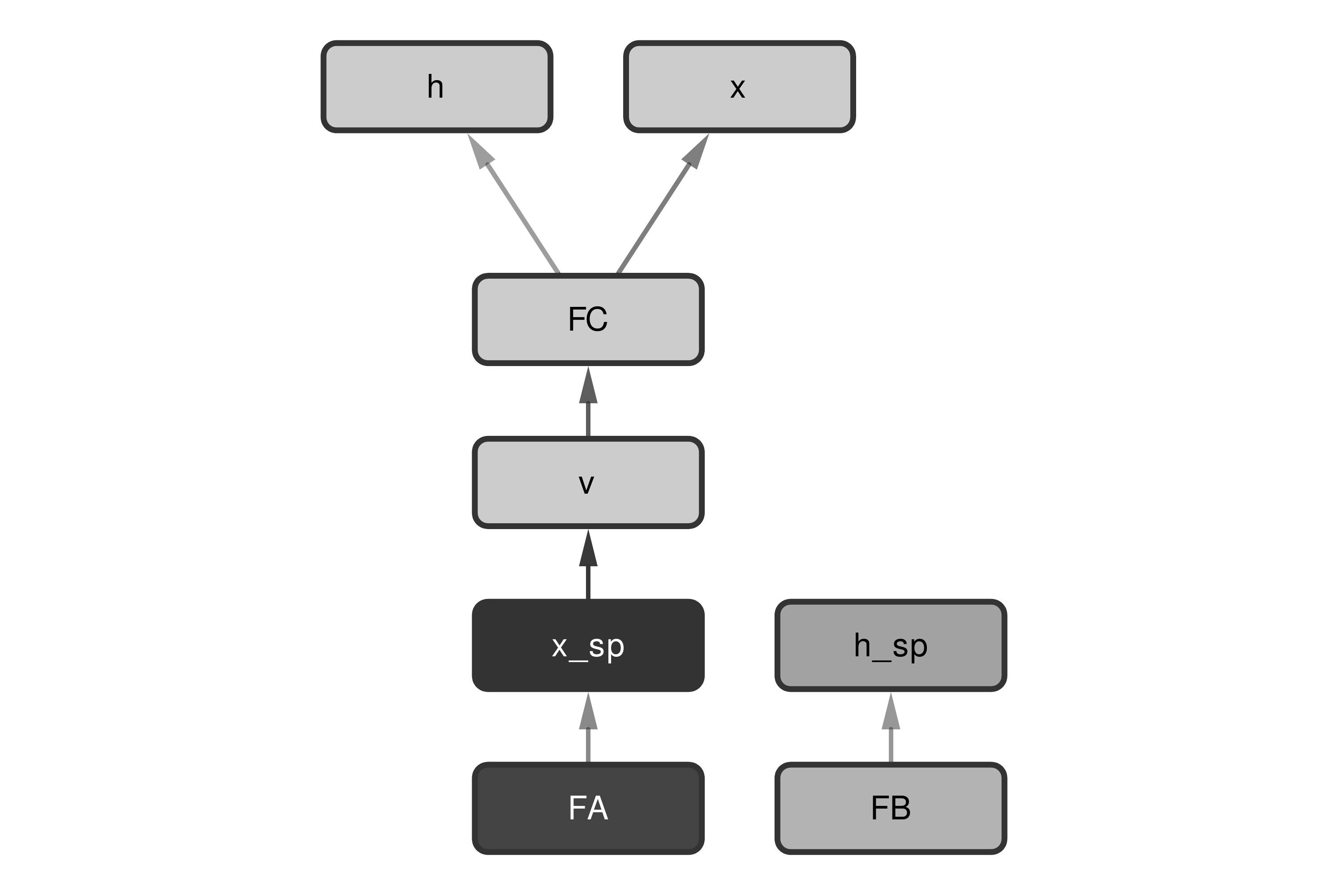}} \\
	\end{tabular}
	
	\caption{Reduced edge ITNs for $x_{\mathrm{sp}}$ disturbance [significance tested]}
	\label{fig:level_concentration_xsp_dist_nosigtest_absolute_ITNs}
\end{figure}

\subsection{Tennessee Eastman Challenge Problem}

The Tennessee Eastman challenge problem \cite{Downs1993} is a popular \ac{FDD} method benchmark \cite{Su2017a} simulation that comes with 20 pre-defined disturbances that can be activated.
Some base layer control strategies have been suggested in literature; the scheme presented by Ricker \cite{Ricker1996} is used as the basis for all tests presented here.

To test the \ac{MTR} approach and ability to handle multiple simultaneous effects two of the random variation disturbances were activated for specific periods of time with an overlapping window.
A random variation in the reactor cooling water (CW) inlet temperature was introduced at a simulation time of $T=37$ hours and deactivated at $T=145$, and random variations in the composition of feed stream E was introduced at $T=109$ and deactivated at $T=217$.
252 hours of operation were simulated at an interval of $5\text{\sc{e}-}4$ hours, capturing a total of 504\ 000 data points for each variable.

\begin{figure}[htbp]
	\centering
	\includegraphics[height=20cm]{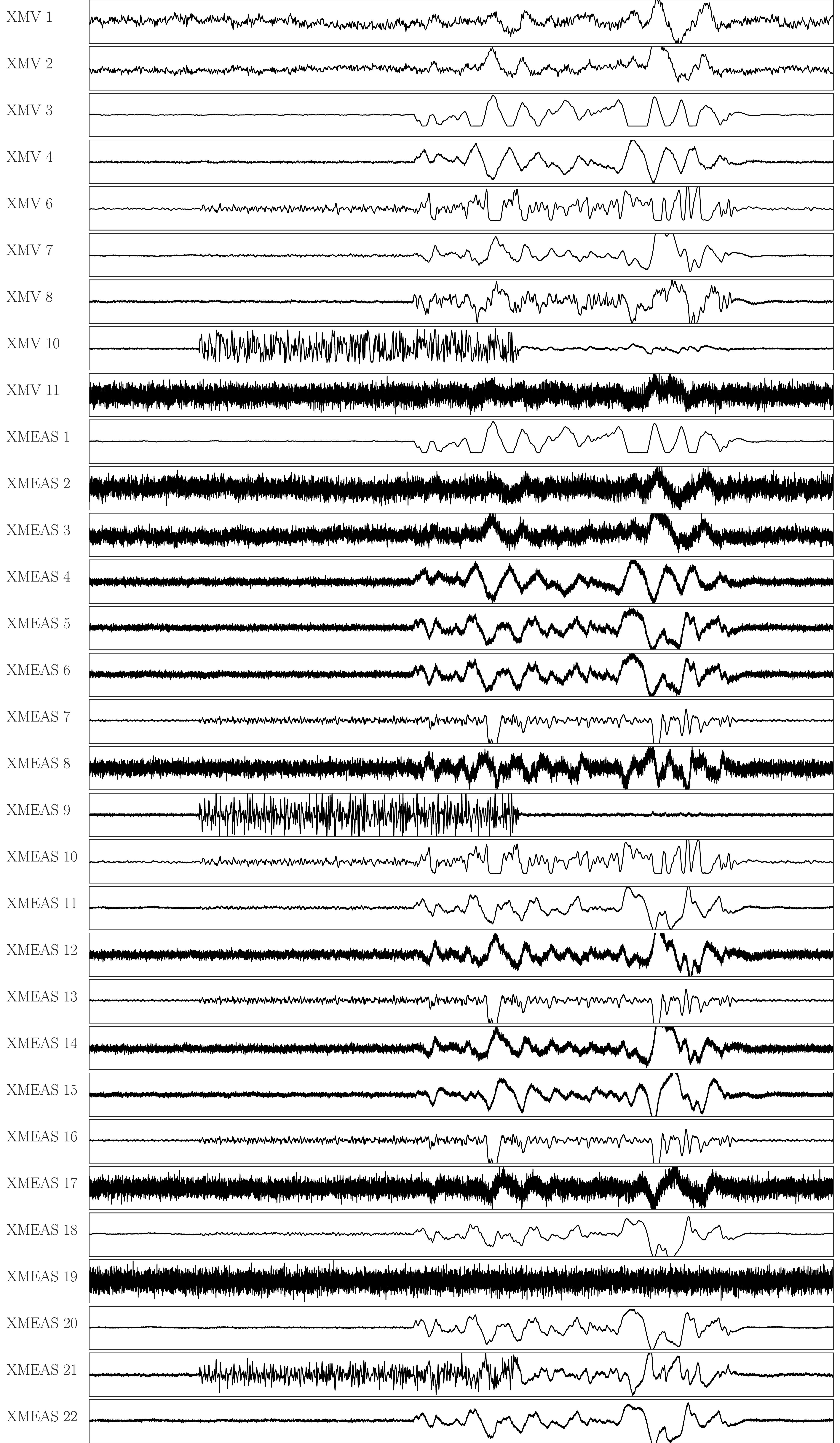}
	
	\caption{High density time series plot of disturbances [standardised]}
	\label{fig:tennessee_eastman_dist_hdts_standardised}
\end{figure}

Figure~\ref{fig:tennessee_eastman_dist_hdts_standardised} presents standardised high density time series plots of all relevant signals generated in the simulation.
Note the overlapping disturbances by comparing XMV 10 with XMV 8, for example.

\clearpage
As discussed in Section~\ref{parameter_selection}, a balance between sampling rate and sample size must be achieved with respect to the process time constants of interest.
The original data was subsampled by a factor of 30 giving a final sampling interval of 0.015 hours or 54 seconds.
At this resolution, if the suggested number of samples of 2\,000 is used, each analysis period covers 30 hours which should be enough to capture the slow dynamics of this process.

In order to achieve a 75\% overlap between consecutive time windows, 28 periods covering 36 hours each was evenly over the 252 hours of simulated operation.
The 36 hours covered resolves to 2\,400 samples per time window.
It was decided to search for the optimal delay over 100 samples or 1.5 hours in each direction, resulting in a final sample size of 2\,200.

The reactor cooling water inlet temperature disturbance was active in isolation during bins 2-10, while the feed composition disturbance was active in isolation for 19-27, with both disturbances present in bins 11-18.
To demonstrate the nature of results obtained, the two disturbances are investigated in isolation as and in an overlapping time window.
It was selected to use the bins where each disturbance has been active for the longest in isolation over the full duration of the bin, that is, bins 10 and 23 for the reactor cooling water and feed composition disturbances respectively.

\begin{figure}[htbp]
	\centering
	\includegraphics[width=\textwidth]{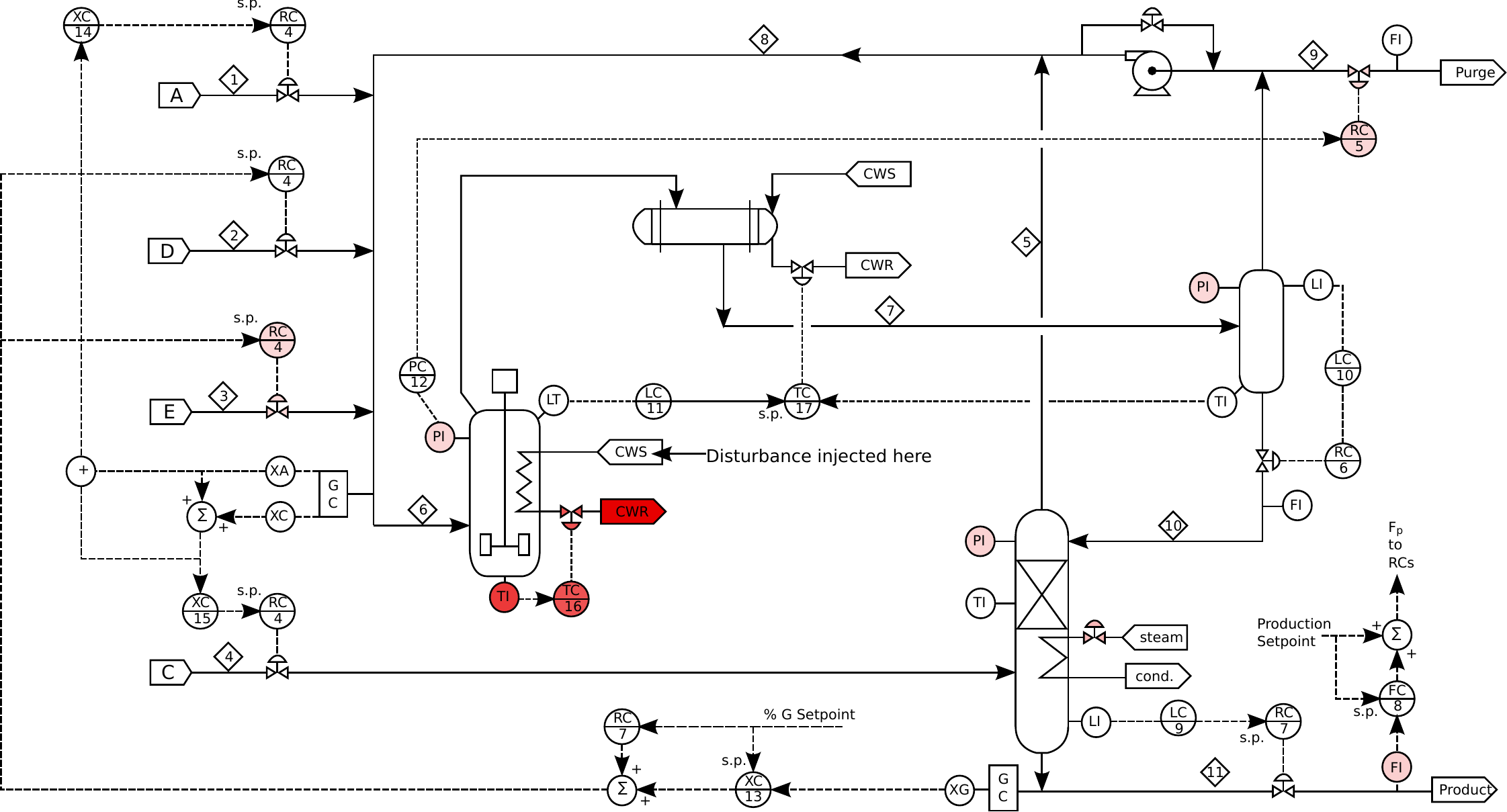}
	\caption{Tennessee Eastman schematic with cooling water inlet temperature disturbance effects highlighted}
	\label{fig:te_bin10_schematic}
	
\end{figure}

To investigate the effects of a reactor cooling water inlet temperature disturbance in isolation, an analysis of the results obtained for bin 10 is presented.
At the start of this bin, the disturbance was already active for 35 hours and can be expected to have propagated through process units with long time constants.
A list relative tag rankings are presented in Table~\ref{tab:te_bin10_noderankings} and the tag closest to the actual source of the disturbance is given the highest importance score by a significant margin, followed by the reactor temperature cooling water flow, which is connected to the reactor temperature via a control loop as can be seen in the schematic Figure~\ref{fig:te_bin10_schematic}. 

By looking at the network arranged in a hierarchical layout, a very clear picture of the situation is obtained.
The reactor cooling water outlet temperature is not receiving any incoming connections indicating that it is an independent source of information flow in the system.
A clear propagation path from the cooling water outlet temperature to reactor temperature and the cooling water flow \ac{MV} is indicated.
Note all the knock-on effects of the cooling water disturbance indicated by the \ac{KSG} estimator, including a number of feedback loops.

\medskip

\begin{minipage}{\linewidth}
	\centering
	\label{tab:te_bin10_noderankings}
	\captionof{table}{Transfer entropy ranking results for CW inlet temperature disturbance [kernel estimator; directional weights; forward delays; significance tested]}
	\begin{tabular}{l l l l l}
		\toprule
		Rank & Tag name & Tag description & Tag type & Relative score \\
		\midrule
		\textcolor{red}{\textbf{1}} & XMEAS 21 & \textcolor{red}{\textbf{Reactor CW outlet temperature}} & \textcolor{red}{\textbf{PV}} & \textcolor{red}{\textbf{\rule{25.00mm}{3mm}}} \\
		\textcolor{red}{2} & XMEAS 9 & \textcolor{red}{Reactor temperature} & \textcolor{red}{CV} & \textcolor{red}{\rule{19.36mm}{3mm}} \\
		\textcolor{red}{3} &  XMV 10 & \textcolor{red}{Reactor CW flow} & \textcolor{red}{MV} & \textcolor{red}{\rule{16.96mm}{3mm}} \\
		4 & XMEAS 19 & Stripper steam flow & PV & {\rule{6.95mm}{3mm}} \\
		5 & XMEAS 17 & Stripper underflow & CV & {\rule{4.69mm}{3mm}} \\
		6 & XMEAS 16 & Stripper pressure & PV & {\rule{4.38mm}{3mm}} \\
		7 & XMEAS 7 & Reactor pressure & CV & {\rule{4.14mm}{3mm}} \\
		8 & XMV 6 & Purge valve & MV & {\rule{3.97mm}{3mm}} \\
		9 & XMEAS 13 & Product separator pressure & PV & {\rule{3.87mm}{3mm}} \\
		10 & XMV 2 & E feed flow & MV &{\rule{3.83mm}{3mm}} \\
		\bottomrule
	\end{tabular}
\end{minipage}
\medskip

To investigate the effects of a feed composition disturbance in isolation, a list of the top tags for bin 23 is presented in Table~\ref{tab:te_bin23_noderankings}.
The actual source of the disturbance is not immediately apparent. 
It is clear that the control loops that get their set points from the composition controller are very active.
However, note that the stripper temperature and compressor work are some of the measurements closest to the disturbance, while most of the other high-ranking nodes point to a disturbance in the composition.

Node centrality scores might be high for units that serve as distribution points for disturbances, even if they are not powerful sources of disruption themselves.
Stripper steam flow (XMEAS 19) was weakly influenced and influences almost all other nodes which result in an unfortunately high place in the ranking scores.

\medskip
\begin{minipage}{\linewidth}
	\centering
	\label{tab:te_bin23_noderankings}
	\captionof{table}{Transfer entropy ranking results for feed composition disturbance [kernel estimator; directional weights; forward delays; significance tested]}
	\begin{tabular}{l l l l l}
		\toprule
		Rank & Tag name & Tag description & Tag type & Relative score \\
		\midrule
		\textcolor{red}{\textbf{1}} & XMEAS 22 & \textcolor{red}{\textbf{Separator CW outlet temperature}} & \textcolor{red}{\textbf{PV}} & \textcolor{red}{\textbf{\rule{25.00mm}{3mm}}} \\
		\textcolor{red}{2} & XMEAS 19 & \textcolor{red}{Stripper steam flow} & \textcolor{red}{PV} & \textcolor{red}{\rule{20.22mm}{3mm}} \\
		\textcolor{red}{3} &  XMEAS 20 & \textcolor{red}{Compressor work} & \textcolor{red}{PV} & \textcolor{red}{\rule{16.20mm}{3mm}} \\
		4 & XMEAS 13 & Product separator pressure & PV & {\rule{15.07mm}{3mm}} \\
		5 & XMV 1 & D feed flow & MV & {\rule{13.46mm}{3mm}} \\
		6 & XMEAS 21 & Reactor CW outlet temperature & PV & {\rule{12.87mm}{3mm}} \\
		7 & XMV 7 & Separator pot liquid flow & MV & {\rule{12.05mm}{3mm}} \\
		8 & XMEAS 16 & Stripper pressure & PV & {\rule{11.88mm}{3mm}} \\
		9 & XMV 3 & A feed flow & MV & {\rule{11.22mm}{3mm}} \\
		10 & XMEAS 11 & Product separator temperature & CV &{\rule{10.64mm}{3mm}} \\
		\bottomrule
	\end{tabular}
\end{minipage}
\medskip

\begin{figure}[htbp]
	\centering
	\includegraphics[width=\textwidth]{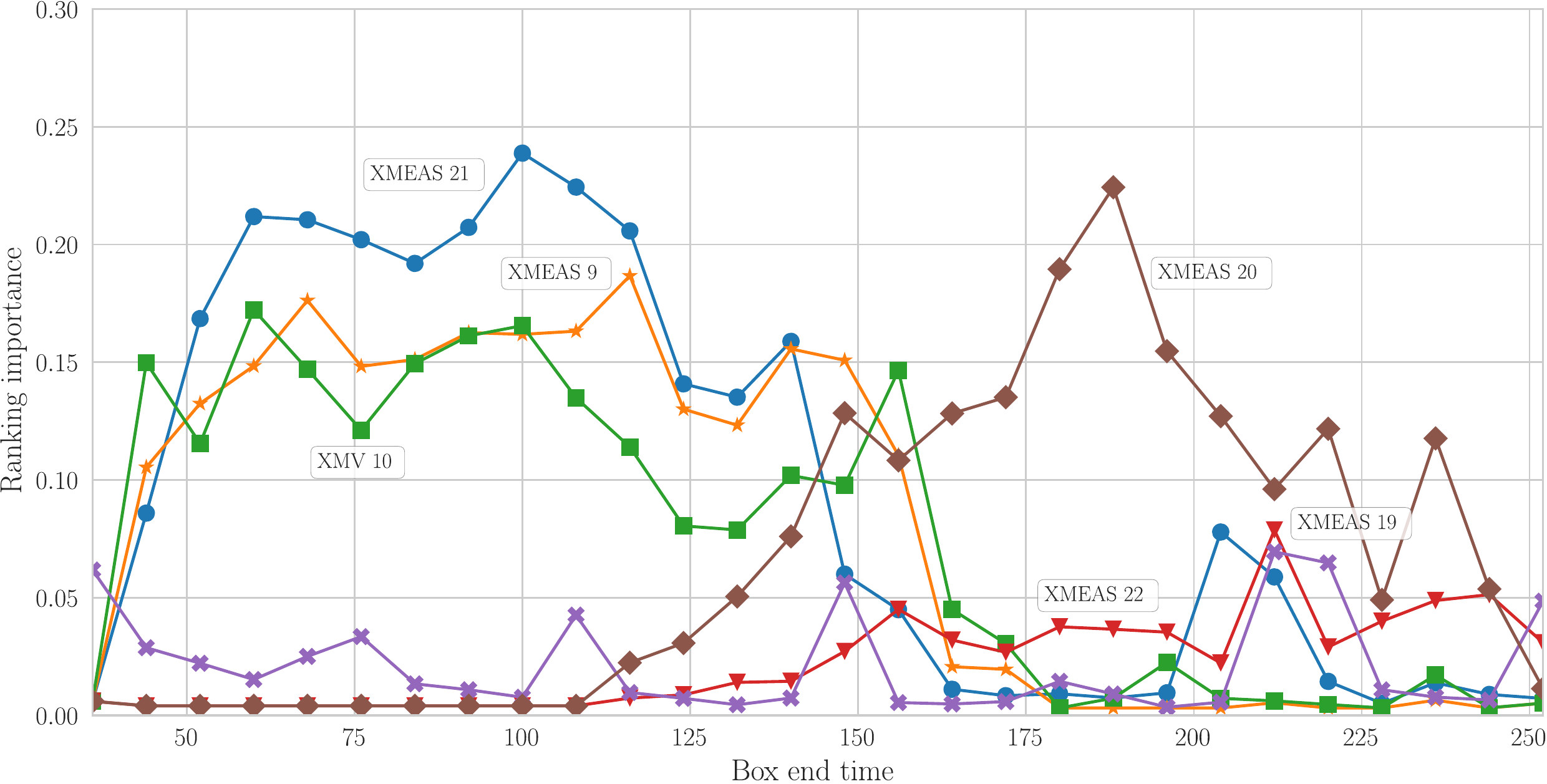}
	\caption{Multiple time region absolute ranking scores for periods covering both reactor CW inlet temperature and stripper feed composition disturbance in Tennessee Eastman process [kernel estimator; directional weights; process limit scaled; forward delays; significance tested]}
	\label{fig:te_dist8_dist11_onedirectional_kernel_walk_absolute}
\end{figure}

Figure~\ref{fig:te_dist8_dist11_onedirectional_kernel_walk_absolute} shows the \ac{MTR} monitoring plot generated over a range spanning both the cooling water and stripper feed composition disturbances, with the most important nodes detected in bins 10 and 23 above indicated.

Note that bin 20 seems to relate more significant disturbance in stripper related variables compared to bin 23 which was analysed in detail.
This is in line with the time series plots in Figure~\ref{fig:tennessee_eastman_dist_hdts_standardised}, and serves to illustrate the potential importance of continuous ranking calculations in facilitating a proper understanding of how a disturbance evolves through a process over time, ultimately allowing for more accurate diagnosis.

\begin{figure}[htbp]
	\centering
	
	\begin{tabular}{cc}
		\subfloat[kernel estimator]{
			\includegraphics[width=0.5\linewidth]{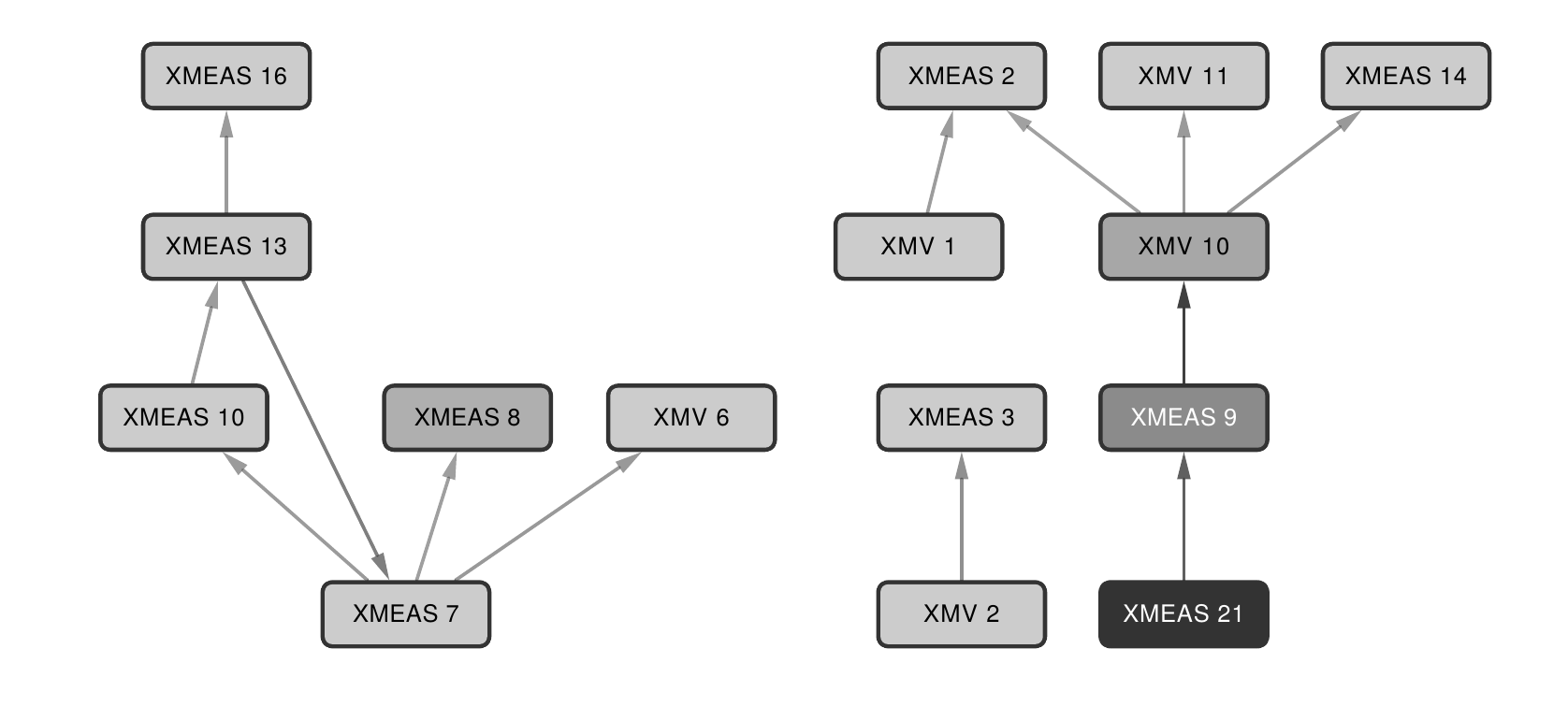}} &
		\subfloat[KSG estimator]{
			\includegraphics[width=0.5\linewidth]{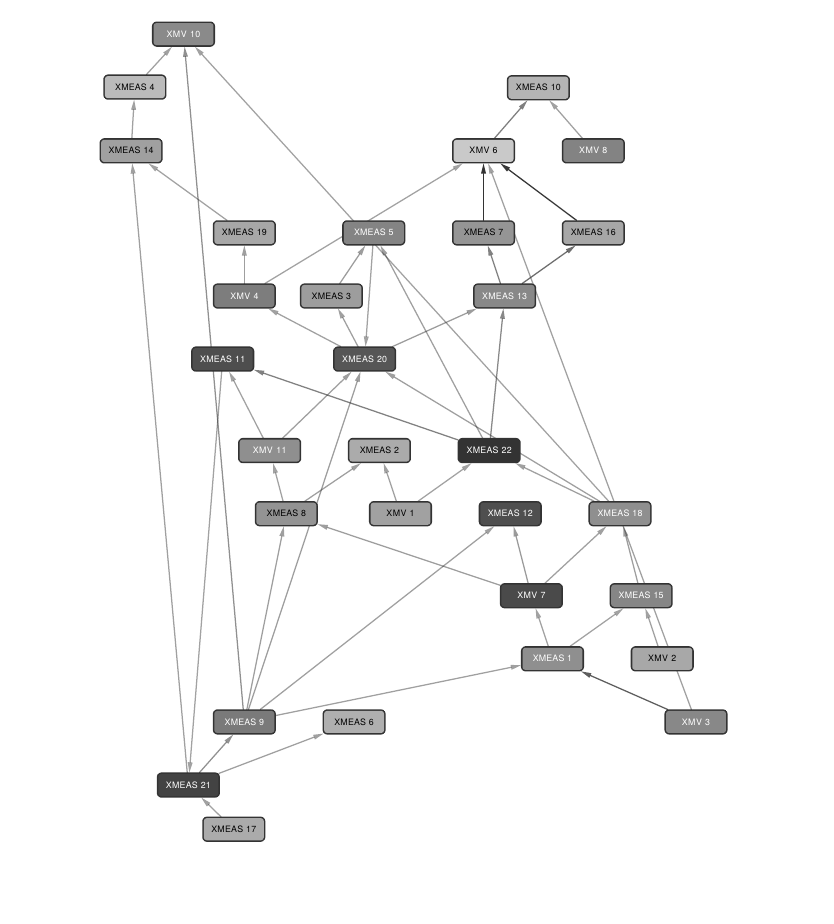}} \\
	\end{tabular}
	
	\caption{Reduced edge ITNs for CW disturbance [simple; bidirectional delays; significance tested]}
	\label{fig:level_concentration_xsp_dist_nosigtest_absolute_ITNs}
\end{figure}

\section{Discussion}

The presented techniques show promise in assisting a process expert to identify the source of a disturbance or fault significantly faster than by following more classical workflows, usually limited to studying time series plots or individual loop \ac{KPI}s.
The methods have proven helpful in two non-trivial industrial fault diagnosis problems encountered in the petrochemical industry and was independently evaluated on test cases \cite{Zalmijn2017}.

Techniques such as auto-embedding and significance testing designed to infer an \ac{ITN} that is more representative of a true causal structure does not assist with obtaining better rankings and may even have a detrimental effect.
When process knowledge based scaling limits are available, the simple kernel transfer entropy method should be used to infer the \ac{ITN}, otherwise directional naive \ac{KSG} estimation should be used.
There is little evidence that embedding improves the ranking results obtained with the \ac{KSG} estimator.
Significance testing is not necessary for getting good \ac{MTR} node ranking plots, and may be safely skipped for fault detection, but should be enabled when the underlying \ac{ITN}s are to be used for fault diagnosis. 

As the results come without robustness guarantees or confidence intervals and may require some effort in interpretation, they are best presented as advisories to panel operators and should not be the basis of automated action pending further development.
The concept of identifying the ``true source'' of a disturbance or fault in a process area that involves multiple interacting elements (especially bi-directional couplings) is ambiguous, since adjusting any number of the elements involved could be likely to improve rejection of the disturbance.
If a ``fault'' due to a poorly tuned controller (as opposed, for example, to a specific faulty valve), then it is likely that re-tuning most of the other controller elements involved can compensate for it.
There are likely to be better and worse solutions, especially with regards to the directionality and potential ill-conditioning of the resulting system.
Finding acceptable tuning parameters might require the use of specialised algorithms combined with expert knowledge and trial-and-error.

As far as could be determined, the following elements is not discussed in prior art related to \ac{TE} based \ac{FDD} on chemical processing plants:
\begin{description}[style=nextline]
	\item[Scaling according to control limits] Results indicate that scaling according to control limits provides significant benefits in certain situations.
	\item[Bi-directional delay analysis with associated significance testing logic]Directional significance testing assists in determining the true direction of interaction between coupled variables.
	This is necessary for an unsupervised optimising search over a range of delays to be robust.
	\item[Use of \ac{KSG} estimator and auto-embedding] The implementation of these techniques in open source code is relatively new and their application in literature mostly limited to neuro- and geoscience.
	The \ac{KSG} estimator exhibits significant advantages in the absence of scaling according to control limits and is also better at accurately detecting long chains of interaction.
\end{description}

\section{Recommended areas for future research}
\subsection{Incorporating meta-data of process elements}
All process states are not equally important from economic or safety perspectives.
Strict control of inventory variables is not required and might even be detrimental to disturbance rejection.
Considering additional information about process elements during the calculation of node importance scores will allow for increased flexibility and relevancy in results.

Biasing the eigenvector network centrality measure with individual weights is discussed in Section~\ref{ranking_bias}, but this might not be sufficient to properly handle multiple sources of potentially conflicting information such as the relative value of material or energy streams, control performance \ac{KPI}s, flags indicating whether a loop is in manual or an element of a \ac{APC} strategy, etc.
The use of ranking algorithms that can optimise a complex objective function \cite{Gao2011} is needed.

\subsection{Cluster ranking}
The probability of indicating a wrong element (false positive) and missing a real source (false negative) is significant when attempting to pinpoint a specific element as the singular source of a fault or disturbance.
A tool tasked with screening faults and disturbances on a plant-wide scale will do well by indicating a cluster of likely elements that can be further analysed with more rigorous information decomposition based techniques.
Chemical plants and their integrated control systems generally behave as dynamic multi-agent processes, and calculating scores for interconnected groups might make analysis more accurate but less precise.

Efficient methods for organising large networks into hierarchical structures with intergroup connections are available \cite{Tsubakino2012}.
In addition, methods that differentiate between inter- and intra-cluster edges, such as the Weighted Inter-Cluster Edge Ranking (WICER) algorithm \cite{Padmanabhan2005} might be useful to consider for this application.

\subsection{Multivariate transfer entropy}
Multivariate transfer entropy can measure directed causal effects between groups of variables and reveal complex synergies \cite{Lizier2011}.
This approach might be useful in the process engineering context on processes that exhibit strong directionality where the impact of relative movement between two or more variables can be more important than the movement of a single variable.

\section{Implementation}
A Python implementation of this method is accessible at \url{https://github.com/SimonStreicher/FaultMap.git}
It relies on the Java Information Dynamics Toolkit (JIDT) \cite{Lizier2008} and NetworkX package \cite{Hagberg2008} for calculation of information-theoretic and network centrality measures, respectively.

\bibliographystyle{unsrt}  
\bibliography{library}

\end{document}